\documentclass[aps,prd,twoside,twocolumn,nofootinbib,10pt,showpacs,floatfix]{revtex4-1}
\usepackage{amsmath,amssymb}
\usepackage{graphicx,bm}
\usepackage{slashed}
\usepackage{epstopdf}
\usepackage{ulem} 
\usepackage[usenames]{color}
\usepackage{float}
\usepackage{hyperref}
\usepackage{subfigure}
\usepackage{subfigure}
\usepackage{rotating}
\usepackage{color}
\usepackage{multirow}
\usepackage{dcolumn}
\usepackage{overpic}
\usepackage{booktabs}
\usepackage{makecell}
\usepackage{diagbox}

\renewcommand\sout{\bgroup \color{red} \ULdepth=-.5ex \ULset}

\newsavebox{\tablebox}
\begin{document}

\title{What can we learn from the electromagnetic properties of hidden-charm molecular pentaquarks with single strangeness?}
\author{Fu-Lai Wang$^{1,2,3}$}
\email{wangfl2016@lzu.edu.cn}
\author{Hong-Yan Zhou$^{1,2}$}
\email{zhouhy20@lzu.edu.cn}
\author{Zhan-Wei Liu$^{1,2,3}$}
\email{liuzhanwei@lzu.edu.cn}
\author{Xiang Liu$^{1,2,3}$\footnote{Corresponding author}}
\email{xiangliu@lzu.edu.cn}
\affiliation{$^1$School of Physical Science and Technology, Lanzhou University, Lanzhou 730000, China\\
$^2$Research Center for Hadron and CSR Physics, Lanzhou University and Institute of Modern Physics of CAS, Lanzhou 730000, China\\
$^3$Lanzhou Center for Theoretical Physics, Key Laboratory of Theoretical Physics of Gansu Province, and Frontiers Science Center for Rare Isotopes, Lanzhou University, Lanzhou 730000, China}

\begin{abstract}
Inspired by the observation of the $P_{cs}(4459)$ and $P_{\psi s}^{\Lambda}(4338)$, we systematically investigate the magnetic moments, the transition magnetic moments, and the radiative decay behaviors of the $S$-wave isoscalar $\Xi_c^{(\prime)}\bar D^{(*)}$ molecular pentaquark states in this work. Our quantitative investigation shows that their electromagnetic properties can provide important hint to decode the inner structure of these discussed isoscalar $\Xi_c^{(\prime)}\bar D^{*}$ molecular pentaquarks. As the potential research issue, we suggest future experiment to focus on the electromagnetic properties of these  isoscalar $\Xi_c^{(\prime)}\bar D^{(*)}$ molecular pentaquarks, which are a typical hidden-charm molecular pentaquark system with single strangeness.
\end{abstract}

\maketitle

\section{Introduction}\label{sec1}

Since the discovery of the $X(3872)$ by the Belle Collaboration in 2003 \cite{Choi:2003ue}, a large number of exotic hadronic states have been observed in the past two decades, which stimulated extensive discussion around them \cite{Liu:2013waa,Hosaka:2016pey,Chen:2016qju,Richard:2016eis,Lebed:2016hpi,Olsen:2017bmm,Guo:2017jvc,Brambilla:2019esw,Liu:2019zoy,Brambilla:2019esw,Chen:2022asf,Meng:2022ozq}. As an important part of the exotic hadron spectroscopy, the molecular state picture was extensively applied to explain them, where we 
have witnessed the big progress on the experimental and theoretical exploration of the hidden-charm molecular pentaquarks \cite{Liu:2013waa,Hosaka:2016pey,Chen:2016qju,Richard:2016eis,Lebed:2016hpi,Olsen:2017bmm,Guo:2017jvc,Brambilla:2019esw,Liu:2019zoy,Chen:2022asf,Meng:2022ozq}. Especially, the observation of three $P_c$ states, $P_c(4312)$, $P_c(4440)$, and $P_c(4457)$, by the LHCb Collaboration in 2019 \cite{Aaij:2019vzc}, provides strong evidence to support the existence of the hidden-charm molecular pentaquarks in the hadron spectroscopy \cite{Wu:2010jy,Wang:2011rga,Yang:2011wz,Wu:2012md,Li:2014gra,Karliner:2015ina,Chen:2015loa}. 

In 2021, LHCb reported the evidence of the  $P_{cs}(4459)$, a candidate of the hidden-charm pentaquark with strangeness, by analyzing the $\Xi_b^-\to J/\psi \Lambda K^-$ process  \cite{LHCb:2020jpq}, which can be viewed as the $\Xi_c \bar{D}^*$ molecular pentaquark state \cite{Hofmann:2005sw,Wu:2010vk,Anisovich:2015zqa,Wang:2015wsa,Feijoo:2015kts,Chen:2015sxa,Chen:2016ryt,Lu:2016roh,Xiao:2019gjd,Shen:2020gpw,Zhang:2020cdi,
Wang:2019nvm,Weng:2019ynv,Chen:2020uif,Peng:2020hql,Chen:2020opr,Liu:2020hcv,Dong:2021juy,Chen:2022onm,Chen:2020kco,Chen:2021cfl,Chen:2021spf,Du:2021bgb,
Hu:2021nvs,Xiao:2021rgp,Zhu:2021lhd}. As indicated by LHCb,  the $P_{cs}(4459)$ state can be as a double-peak structure \cite{LHCb:2020jpq} similar to the case for the $P_c(4450)$ enhancement structure \cite{Aaij:2015tga}, which can be replaced by two substructures $P_c(4440)$ and $P_c(4457)$ \cite{Aaij:2019vzc}. Very recently, LHCb reported the observation of the $P_{\psi s}^\Lambda(4338)$ in the $B^-\to J/\psi\Lambda\bar{p}$ process \cite{Pcs}. Obviously, the observed $P_{cs}(4459)$ and $P_{\psi s}^{\Lambda}(4338)$ not only make the family of the hidden-charm pentaquark with single strangeness becomes abundant, but also inspire theorist's interest in investigating the $\Xi_c\bar{D}^{(*)}$ molecular states \cite{Wang:2022neq,Wang:2022mxy,Karliner:2022erb,Yan:2022wuz,Meng:2022wgl}.

In Ref. \cite{Wang:2022mxy}, we indicated the existence of characteristic spectrum of the $\Xi_c^{(\prime)}\bar D^{(*)}$-type hidden-charm molecular pentaquarks with single strangeness when checking the $S$-wave $\Xi_c^{(\prime)}\bar D^{(*)}$ interactions quantitatively. In fact, this behavior was also found by Karliner and Rosner in Ref. \cite{Karliner:2022erb}. When facing such experimental and theoretical progresses of exploring the hidden-charm molecular pentaquarks with single strangeness, it is natural to expect that the theorists should pay more attention to exploring other properties of the $S$-wave isoscalar $\Xi_c^{(\prime)}\bar D^{(*)}$ molecular states, which is valuable to reveal the mystery behind these novel phenomena. 

As is well known, the study of the electromagnetic properties of the hadrons can provide new insight to reveal their inner structures. In particular, exploring the magnetic moments of hadron can be applied to  distinguish their spin-parity quantum numbers. For example, the magnetic moment of the $\Sigma_c\bar D^{*}$ molecular pentaquark with $I(J^P)=1/2(1/2^-)$ is obviously different from that of the $\Sigma_c\bar D^{*}$ state with $I(J^P)=1/2(3/2^-)$ \cite{Wang:2016dzu,Li:2021ryu}. This observation can be applied to clarify the spin-parity quantum numbers of the $P_c(4440)$ and $P_c(4457)$ \cite{Aaij:2019vzc} \cite{Wu:2010jy,Wang:2011rga,Yang:2011wz,Wu:2012md,Li:2014gra,Karliner:2015ina,Chen:2015loa}. In the past several decades, 
different theoretical methods or approaches were proposed
to investigate the electromagnetic properties of the hadronic states \cite{Meng:2022ozq}. Among them, the constituent quark model is a popular way to study the magnetic moments of the decuplet and octet baryons quantitatively \cite{Schlumpf:1993rm,Ramalho:2009gk}.

Along this line, for further presenting the inner structures of the $S$-wave isoscalar $\Xi_c^{(\prime)}\bar D^{(*)}$ molecular states, it is essential to obtain the information of their magnetic moments, transition magnetic moments, and radiative decay behaviors, which will be the main task of this work. For achieving this goal, we adopt the constituent quark model in our realistic calculation since it has been  adopted to investigate the magnetic moments of the hadronic molecular states \cite{Liu:2003ab,Wang:2016dzu,Deng:2021gnb,Gao:2021hmv,Li:2021ryu,Zhou:2022gra}. In addition, the $S$-$D$ wave mixing effect \cite{Wang:2022mxy} is taken into account in our calculation. By this effort,  we illustrate the electromagnetic properties of the $S$-wave isoscalar $\Xi_c^{(\prime)}\bar D^{(*)}$ molecular pentaquarks, and hope that the present work may inspire our colleagues to further focus on the electromagnetic properties of the $S$-wave isoscalar $\Xi_c^{(\prime)}\bar D^{(*)}$ molecules in the following years. In Ref. \cite{Ozdem:2022kei}, the magnetic moments of the $P_{\psi s}^{\Lambda}(4338)$ and $P_{cs}(4459)$ states were obtained within the hadronic molecular picture by the QCD sum rule.

An outline of this paper is as follows. After Introduction, the detailed deduction of the magnetic moments and the transition magnetic moments related to the $S$-wave charmed baryons $\Xi_c^{(\prime)}$, the $S$-wave anti-charmed mesons $\bar D^{(*)}$, and the $S$-wave isoscalar $\Xi_c^{(\prime)}\bar D^{(*)}$ molecular states will be given in Sec. \ref{sec2}. With this preparation, we present the numerical results and the corresponding discussion of the magnetic moments, the transition magnetic moments, and the radiative decay behaviors of the $S$-wave isoscalar $\Xi_c^{(\prime)}\bar D^{(*)}$ molecular pentaquarks in Sec. \ref{sec3}. Finally, this work ends with a short summary in Sec. \ref{sec4}.

\section{The electromagnetic properties of the  hidden-charm molecular pentaquarks with single strangeness}\label{sec2}

In this section, the main task is to deduce the magnetic moments and the transition magnetic moments of the $S$-wave charmed baryons $\Xi_c^{(\prime)}$, the $S$-wave anti-charmed mesons $\bar D^{(*)}$, and the $S$-wave isoscalar $\Xi_c^{(\prime)}\bar D^{(*)}$ molecular pentaquarks, where we adopt the constituent quark model in the calculation. We should emphasize that the constituent quark model was extensively applied to study various properties of the hadronic states in the past decades \cite{Liu:2013waa,Hosaka:2016pey,Chen:2016qju,Richard:2016eis,Lebed:2016hpi,Brambilla:2019esw,Liu:2019zoy,Chen:2022asf,Olsen:2017bmm,Guo:2017jvc,Meng:2022ozq}, where the magnetic moments of the hadronic molecular states were focused \cite{Liu:2003ab,Wang:2016dzu,Deng:2021gnb,Gao:2021hmv,Li:2021ryu,Zhou:2022gra}.


In the present work, the adopted model and convention of getting the hadronic magnetic moments and the transition magnetic moments are the same as those given in Refs.  \cite{Li:2021ryu,Zhou:2022gra}. Within the constituent quark model, the magnetic moment of the hadronic state ${\mu}$ can be written as the sum of the spin magnetic moment ${\mu}_{{\rm spin}}$ and the orbital magnetic moment ${\mu}_{{\rm orbital}}$ from its constituents \cite{Huang:2004tn,Wang:2016dzu,Liu:2003ab,Zhou:2022gra}, i.e.,
\begin{eqnarray}
 {\mu}&=&{\mu}_{{\rm spin}}+{\mu}_{{\rm orbital}}.
\end{eqnarray}
Here, the spin magnetic moment ${\mu}_{{\rm spin}}$ and the orbital magnetic moment ${\mu}_{{\rm orbital}}$ can be related to the spin of each constituent and the orbital angular momenta between its constituents, respectively. Subsequently, we define the spin magnetic moment ${\mu}_{{\rm spin}}$ and the orbital magnetic moment ${\mu}_{{\rm orbital}}$ adopted in the present work.

In practice, the magnetic moment $\mu_{H_0}$ and the transition magnetic moment $\mu_{H_1 \to H_2}$ can be calculated by the $z$-component of the magnetic moment operator $\hat{\mu}_z$ sandwiched by the corresponding wave functions of the investigated hadrons, and the general expression is
\begin{eqnarray}
\mu_{H_0}&=&\left\langle{H_0} \left|\hat{\mu}_z \right| {H_0}\right\rangle,\\
\mu_{H_1 \to H_2}&=&\left\langle{H_2} \left|\hat{\mu}_z \right| {H_1}\right\rangle, \label{expectationvalue}
\end{eqnarray}
where $|{H_i}\rangle$ stands for the corresponding wave function of the investigated hadronic state. Thus, the first task is to calculate the matrix elements $\left\langle{H_0} \left|\hat{\mu}_z \right| {H_0}\right\rangle$ and $\left\langle{H_2} \left|\hat{\mu}_z \right| {H_1}\right\rangle$ for  extracting the hadronic magnetic moment and the transition magnetic moment.

The wave function of the hadronic state $\psi$ is composed of the color wave function $\omega_{\rm{color}}$, the flavor wave function $\chi_{\rm{flavor}}$, the spin wave function $\chi_{\rm{spin}}$, and the spatial wave function $R_{\rm{space}}$, which can be factorized as
\begin{equation}
\psi=\omega_{\rm{color}}\otimes\chi_{\rm{flavor}}\otimes\chi_{\rm{spin}}\otimes R_{\rm{space}}.
\end{equation}
For the hadronic wave function, we should emphasize:
\begin{enumerate}
  \item The color wave functions do not affect the magnetic moments of hadrons, and we do not need to consider them explicitly;
  \item For the flavor-spin wave function, we need to consider the requirement of the symmetry, which is an important step when getting the magnetic moment and the transition magnetic moment of the hadronic state;
  \item For the spatial wave function \cite{Zhou:2022gra}, we usually do not mention it when calculating the magnetic moment and the transition magnetic moment of the hadronic state, since the $S$ waves do not contribute to the magnetic moments. However, if considering the $S$-$D$ wave mixing effect, we 
 should take the spatial wave functions of these involved mixing channels as input, which can be numerically obtained from solving the Schr{\"o}dinger equation for the mass spectrum.
\end{enumerate}


Due to the absence of experimental data of the magnetic moments and the transition magnetic moments of the $S$-wave charmed baryons $\Xi_c^{(\prime)}$ and the $S$-wave anti-charmed mesons $\bar D^{(*)}$, we first calculate them within the constituent quark model. For $\Xi_c^{(\prime)}$ and $\bar D^{(*)}$, there only exist the spin magnetic moments $\vec{\mu}_{{\rm spin}}$ since the orbital angular momentum among quarks is 0, and at quark level \cite{Huang:2004tn,Wang:2016dzu,Liu:2003ab,Zhou:2022gra}
\begin{eqnarray}
 \vec{\mu}_{{\rm spin}}&=&\sum_{i}\frac{Q_i}{2M_i}\overrightarrow{\sigma_{i}},
\end{eqnarray}
where $Q_i$, $M_i$, and $\overrightarrow{\sigma_{i}}$ denote the charge, mass, and Pauli's spin matrix of the $i$-th quark, respectively.

In order to obtain the magnetic moments and the transition magnetic moments of $\Xi_c^{(\prime)}$ and $\bar D^{(*)}$, we need to construct their flavor wave functions and spin wave functions. Based on the flavor symmetries of the light diquark, the charmed baryons can be categorized into the $\bar 3_F$ and $6_F$ flavor representations, which correspond to the flavor antisymmetry and symmetry for the light diquark, respectively. Here, $\Xi_c$ with $J^P=1/2^+$ denotes the $S$-wave charmed baryon in the $\bar{3}_F$ flavor representation, while $\Xi_c^\prime$ with $J^P=1/2^+$ is for the $6_F$ flavor representation. In Table \ref{Twavefunctions}, we collect these flavor wave functions $\chi_{\rm{flavor}}$ and the spin wave functions $\chi_{\rm{spin}}$ of $\Xi_c^{(\prime)}$ and $\bar D^{(*)}$ \cite{Simonis:2016pnh,Workman:2022ynf}, which are adopted to calculate the matrix elements $\mu_{H_0}=\left\langle{H_0} \left|\hat{\mu}_z \right| {H_0}\right\rangle$ and
$\mu_{H_1 \to H_2}=\left\langle{H_2} \left|\hat{\mu}_z \right| {H_1}\right\rangle$.
\renewcommand\tabcolsep{0.13cm}
\renewcommand{\arraystretch}{1.50}
\begin{table}[!htbp]
\caption{The flavor wave functions $\chi_{\rm{flavor}}$ and the spin wave functions $\chi_{\rm{spin}}$ of the $S$-wave charmed baryons $\Xi_c^{(\prime)}$ and the $S$-wave anti-charmed mesons $\bar D^{(*)}$. Here, $S$ and $S_3$ are the spin and its third component of the discussed hadron, while the arrow denotes the third component of the quark spin. }
\label{Twavefunctions}
\begin{tabular}{c|l|l}
\toprule[1.0pt]\toprule[1.0pt]
States&$\left|S,S_3\right\rangle$ & $\chi_{\rm{flavor}}\otimes\chi_{\rm{spin}}$ \\\hline
\multirow{2}{*}{$\Xi_c^{+}$}&$\left|\frac{1}{2},\frac{1}{2}\right\rangle$ & $\frac{1}{\sqrt{2}}\left(usc-suc\right)\otimes\frac{1}{\sqrt{2}}\left(\uparrow\downarrow\uparrow-\downarrow\uparrow\uparrow\right)$ \\
                     &$\left|\frac{1}{2},-\frac{1}{2}\right\rangle$ & $\frac{1}{\sqrt{2}}\left(usc-suc\right)\otimes\frac{1}{\sqrt{2}}\left(\uparrow\downarrow\downarrow-\downarrow\uparrow\downarrow\right)$ \\
\multirow{2}{*}{$\Xi_c^{0}$}&$\left|\frac{1}{2},\frac{1}{2}\right\rangle$ & $\frac{1}{\sqrt{2}}\left(dsc-sdc\right)\otimes\frac{1}{\sqrt{2}}\left(\uparrow\downarrow\uparrow-\downarrow\uparrow\uparrow\right)$ \\
                   &$\left|\frac{1}{2},-\frac{1}{2}\right\rangle$ & $\frac{1}{\sqrt{2}}\left(dsc-sdc\right)\otimes\frac{1}{\sqrt{2}}\left(\uparrow\downarrow\downarrow-\downarrow\uparrow\downarrow\right)$ \\
\multirow{2}{*}{$\Xi_c^{\prime+}$}&$\left|\frac{1}{2},\frac{1}{2}\right\rangle$ & $\frac{1}{\sqrt{2}}\left(usc+suc\right)\otimes\frac{1}{\sqrt{6}}\left(2\uparrow\uparrow\downarrow-\downarrow\uparrow\uparrow-\uparrow\downarrow\uparrow\right)$ \\
           &$\left|\frac{1}{2},-\frac{1}{2}\right\rangle$ & $\frac{1}{\sqrt{2}}\left(usc+suc\right)\otimes\frac{1}{\sqrt{6}}\left(\downarrow\uparrow\downarrow+\uparrow\downarrow\downarrow-2\downarrow\downarrow\uparrow\right)$ \\
\multirow{2}{*}{$\Xi_c^{\prime0}$}&$\left|\frac{1}{2},\frac{1}{2}\right\rangle$ & $\frac{1}{\sqrt{2}}\left(dsc+sdc\right)\otimes\frac{1}{\sqrt{6}}\left(2\uparrow\uparrow\downarrow-\downarrow\uparrow\uparrow-\uparrow\downarrow\uparrow\right)$ \\
              &$\left|\frac{1}{2},-\frac{1}{2}\right\rangle$ & $\frac{1}{\sqrt{2}}\left(dsc+sdc\right)\otimes\frac{1}{\sqrt{6}}\left(\downarrow\uparrow\downarrow+\uparrow\downarrow\downarrow-2\downarrow\downarrow\uparrow\right)$ \\
$\bar D^{0}$&$\left|0,0\right\rangle$ & $\bar c u\otimes\frac{1}{\sqrt{2}}\left(\uparrow\downarrow-\downarrow\uparrow\right)$ \\
$D^{-}$&$\left|0,0\right\rangle$ & $\bar c d\otimes\frac{1}{\sqrt{2}}\left(\uparrow\downarrow-\downarrow\uparrow\right)$ \\
\multirow{3}{*}{$\bar D^{*0}$}&$\left|1,1\right\rangle$ & $\bar c u\otimes\uparrow\uparrow$ \\
         &$\left|0,0\right\rangle$ & $\bar c u\otimes\frac{1}{\sqrt{2}}\left(\uparrow\downarrow+\downarrow\uparrow\right)$ \\
         &$\left|1,-1\right\rangle$ & $\bar c u\otimes\downarrow\downarrow$ \\
\multirow{3}{*}{$D^{*-}$}&$\left|1,1\right\rangle$ & $\bar c d\otimes\uparrow\uparrow$ \\
          &$\left|0,0\right\rangle$ & $\bar c d\otimes\frac{1}{\sqrt{2}}\left(\uparrow\downarrow+\downarrow\uparrow\right)$ \\
           &$\left|1,-1\right\rangle$ & $\bar c d\otimes\downarrow\downarrow$ \\
\bottomrule[1.0pt]\bottomrule[1.0pt]
\end{tabular}
\end{table}

For these $\Xi_c^{(\prime)}$ and $\bar D^{(*)}$, their magnetic moments can be estimated by the $z$-component of the magnetic moment operator $\hat{\mu}_z$ sandwiched by the corresponding flavor-spin wave functions of $\Xi_c^{(\prime)}$ and $\bar D^{(*)}$, respectively. Here, we need to indicate that the magnetic moment of the quark can be obtained by the following matrix elements
\begin{eqnarray}
  \left\langle q\uparrow \left| \hat{\mu}_z \right| q\uparrow \right\rangle &=& \frac{Q_q}{2M_q},\\
  \left\langle q\downarrow  \left| \hat{\mu}_z \right| q\downarrow  \right\rangle &=& - \frac{Q_q}{2M_q}.
\end{eqnarray}
Here, the arrow stands for the third component of the quark spin, while $Q_q$ and $M_q$ are the quark charge and mass, respectively. For example,
\begin{eqnarray}
\mu_{\bar D^{*0}}&=&\left\langle \chi_{\bar D^{*0}}^{S=1;\,S_3=1} \left|\hat{\mu}_z\right| \chi_{\bar D^{*0}}^{S=1;\,S_3=1} \right\rangle\nonumber\\
&=& \left\langle \bar c u\uparrow\uparrow \left|\hat{\mu}_z\right| \bar c u\uparrow\uparrow\right\rangle\nonumber\\
&=&\mu_{\bar c}+\mu_{u}.
\end{eqnarray}
Thus, the expression of the magnetic moment of the $S$-wave charmed meson $\bar D^{*0}$ is $\mu_{\bar c}+\mu_{u}$. In this work, we adopt the constituent quark masses $m_{u}=0.336\,\mathrm{GeV}$, $m_{d}=0.336\,\mathrm{GeV}$, $m_{s}=0.450\,\mathrm{GeV}$, and $m_{c}=1.680\,\mathrm{GeV}$ \cite{Kumar:2005ei} to present the hadronic magnetic moments and transition magnetic moments quantitatively, which are widely used to study the hadronic magnetic moments \cite{Li:2021ryu,Zhou:2022gra}. By numerical calculation, we can obtain the magnetic moment of $\bar D^{*0}$ is $1.489\mu_{N}$. Here, $\mu_N={e}/{2m_N}$ is the nuclear magnetic moment with $m_{N}=938\,\mathrm{MeV}$ as the nuclear mass \cite{Workman:2022ynf}, which is the unit of the magnetic moment.

In Table \ref{MT}, we list the expressions and numerical results of the magnetic moments of $\Xi_c^{(\prime)}$ and $\bar D^{(*)}$, which are expressed as the combination of the magnetic moments of their constituent quarks. In addition, we compare our obtained numerical results with those from other theoretical works, and we can see they are consistent \cite{Kumar:2005ei,Faessler:2006ft,Glozman:1995xy,Simonis:2018rld,Aliev:2008ay,Aliev:2015axa,Wang:2019mhm,Bose:1980vy}.
\renewcommand\tabcolsep{0.09cm}
\renewcommand{\arraystretch}{1.50}
\begin{table}[!htbp]
  \caption{The magnetic moments of the $S$-wave charmed baryons $\Xi_c^{(\prime)}$ and the $S$-wave anti-charmed mesons $\bar D^{(*)}$. Here, the magnetic moments of the $S$-wave anti-charmed mesons $\bar{D}^{0}$ and $D^{-}$ are zero, the magnetic moment is in unit of the nuclear magnetic moment $\mu_N$, and $\mu_q={Q_q}/{2M_q}$ with $Q_q$ and $M_q$ are the quark charge and mass, respectively.}
  \label{MT}
\begin{tabular}{c|cc|cc}
\toprule[1.0pt]\toprule[1.0pt]
{Hadrons} &  Expressions & Results & \multicolumn{2}{c}{Other works} \\ \cline{1-5}
$\Xi^{+}_c$ & $\mu_c$ & $0.372$ & 0.37 \cite{Kumar:2005ei}&  0.37 \cite{Faessler:2006ft} \\
$\Xi^{0}_c$  & $\mu_c$ & $0.372$ & 0.366 \cite{Kumar:2005ei}&  0.38 \cite{Glozman:1995xy} \\
$\Xi^{\prime +}_c$ & $\frac{2}{3}\mu_u+\frac{2}{3}\mu_s-\frac{1}{3}\mu_c$ & $0.654$ & 0.65 \cite{Glozman:1995xy} &0.633 \cite{Simonis:2018rld} \\
$\Xi^{\prime 0}_c$ & $\frac{2}{3}\mu_d+\frac{2}{3}\mu_s-\frac{1}{3}\mu_c$ & $-1.208$ &$-1.23$ \cite{Aliev:2008ay}&$-1.23$ \cite{Aliev:2015axa}\\
$\bar{D}^{*0}$ & $\mu_{\overline{c}}+\mu_u$ & $1.489$ & 1.28 \cite{Simonis:2018rld}& 1.48 \cite{Wang:2019mhm}\\
$D^{*-}$ & $\mu_{\overline{c}}+\mu_d$ & $-1.303$ & $-1.31$ \cite{Wang:2019mhm} &$-1.17$ \cite{Bose:1980vy}  \\
\bottomrule[1.0pt]\bottomrule[1.0pt]
\end{tabular}
\end{table}

Additionally, the transition magnetic moments of $\Xi_c^{(\prime)}$ and $\bar D^{(*)}$ can be obtained by calculating the matrix element $\mu_{H_1 \to H_2}=\left\langle{H_2} \left|\hat{\mu}_z \right| {H_1}\right\rangle$, and the initial and final states take the same third components of the spin when deducing the hadronic transition magnetic moment. For instance,
\begin{eqnarray}
  \mu_{\bar D^{*0}\rightarrow \bar D^0}&=&\left\langle\chi_{\bar D^{0}}^{S=0;\,S_3=0} \left|\hat{\mu}_z\right| \chi_{\bar D^{*0}}^{S=1;\,S_3=0}\right\rangle \notag\\
  &=& \left\langle \frac{\bar c u\uparrow \downarrow-\bar c u\downarrow\uparrow}{\sqrt{2}}\right|\hat{\mu}_z\left|\frac{\bar c u\uparrow\downarrow+\bar c u\downarrow\uparrow}{\sqrt{2}} \right\rangle \notag\\
  &=&\mu_{\bar c}-\mu_{u}.
\end{eqnarray}
Thus, the expression and numerical result of the transition magnetic moment of the $\bar D^{*0}\to \bar D^0\gamma$ process are $\mu_{\bar c}-\mu_{u}$ and $-2.234\mu_{N}$, respectively.

In Table \ref{TMT}, we present the expressions and numerical results of the transition magnetic moments of $\Xi_c^{(\prime)}$ and $\bar D^{(*)}$. In order to check the reliability of these obtained transition magnetic moments, we compare our obtained numerical results with those from other theoretical models, and our results are close to other theoretical predictions \cite{Kumar:2005ei,Hazra:2021lpa,Simonis:2018rld,Wang:2019mhm}.
\renewcommand\tabcolsep{0.06cm}
\renewcommand{\arraystretch}{1.50}
\begin{table}[!htbp]
  \caption{The transition magnetic moments of the $S$-wave charmed baryons $\Xi_c^{(\prime)}$ and the $S$-wave anti-charmed mesons $\bar D^{(*)}$. Here, the transition magnetic moment is in unit of the nuclear magnetic moment $\mu_N$.}
  \label{TMT}
\begin{tabular}{l|cc|cc}
\toprule[1.0pt]\toprule[1.0pt]
{Decay modes} & Expressions & Results & \multicolumn{2}{c}{Other works} \\ \cline{1-5}
$\Xi^{\prime+}_c \to \Xi^{+}_c\gamma$ & $\frac{1}{\sqrt{3}}(\mu_s-\mu_u)$ & $-1.476$ & $-1.39$ \cite{Kumar:2005ei}&$-1.4282$ \cite{Hazra:2021lpa}\\
$\Xi^{\prime0}_c \to \Xi^{0}_c\gamma$ & $\frac{1}{\sqrt{3}}(\mu_s-\mu_d)$ & $0.136$ & 0.13 \cite{Kumar:2005ei}& 0.138 \cite{Simonis:2018rld}   \\
$\bar{D}^{*0} \to \bar{D}^{0}\gamma$ & $\mu_{\overline{c}}-\mu_u$ & $-2.234$ & $-2.13$ \cite{Wang:2019mhm}&   \\
$D^{*-} \to D^{-}\gamma$ & $\mu_{\overline{c}}-\mu_d$ & $0.558$ & 0.54 \cite{Wang:2019mhm}&   \\
\bottomrule[1.0pt]\bottomrule[1.0pt]
\end{tabular}
\end{table}


In our previous work \cite{Wang:2022mxy}, we already studied the $S$-wave $\Xi_c^{(\prime)}\bar D^{(*)}$-type mass spectrum of the hidden-charm molecular pentaquarks with strangeness, and our numerical results suggest that all of the $S$-wave isoscalar $\Xi_c^{(\prime)}\bar D^{(*)}$ states can be recommended as the hidden-charm molecular pentaquark candidates with strangeness. In this subsection, we mainly illustrate how to deduce the magnetic moments, the transition magnetic moments, and the radiative decay behaviors of these hidden-charm molecular pentaquark candidates with strangeness.

First we assume these molecular states are in $S$ wave, and then we further take into account the $S$-$D$ wave mixing effect. The $S$-wave hadronic state only has the spin magnetic moment ${\mu}_{{\rm spin}}$, and the $D$-wave hadronic state contains the spin magnetic moment ${\mu}_{{\rm spin}}$ and the orbital magnetic moment ${\mu}_{{\rm orbital}}$ from its constituents. Here, the orbital magnetic moments ${\mu}_{{\rm orbital}}$ of the $D$-wave $\Xi_c^{(\prime)}\bar D^{(*)}$ channels can be written as \cite{Huang:2004tn,Wang:2016dzu,Liu:2003ab,Zhou:2022gra,Girdhar:2015gsa,Li:2021ryu}
\begin{eqnarray}
 \vec{\mu}_{{\rm orbital}}&=&\mu_{mb}^L\vec{L}=\frac{M_{m}}{M_{b}+M_{m}}\frac{Q_b}{2M_b}\vec{L}+\frac{M_{b}}{M_{b}+M_{m}}\frac{Q_m}{2M_m}\vec{L},\nonumber\\
\end{eqnarray}
where the subscripts $b$ and $m$ correspond to the $S$-wave charmed baryons $\Xi_c^{(\prime)}$ and the $S$-wave anti-charmed mesons $\bar D^{(*)}$, respectively, and $\vec{L}$ denotes the orbital angular momenta between $\Xi_c^{(\prime)}$ and $\bar D^{(*)}$. Here, we use the formula ${\hat L}_zY_{L,m_L}=m_LY_{L,m_L}$ and $\int Y_{L,m_L}^{\dag}Y_{L,m_L} {\rm sin} \theta d \theta d \phi=1$ when deducing the hadronic orbital magnetic moments.

\renewcommand\tabcolsep{0.20cm}
\renewcommand{\arraystretch}{1.50}
\begin{table}[!htbp]
\caption{The flavor wave functions $\chi_{\rm{flavor}}$ and the spin wave functions $\chi_{\rm{spin}}$ of the $S$-wave isoscalar $\Xi_c^{(\prime)}\bar D^{(*)}$ systems. Here, $I$ and $I_3$ are the isospin and its third component of the investigated system, respectively, while $S$ and $S_3$ are the spin and its third component of the investigated system, respectively. }
\label{Pwavefunctions}
\begin{tabular}{l|l|l}
\toprule[1.0pt]
\toprule[1.0pt]
Systems& $\left|I,I_3\right\rangle$ & $\chi_{\rm{flavor}}$ \\
\hline
$\Xi_c^{(\prime)}\bar{D}^{(*)}$&$\left|0,0\right\rangle$ & $\frac{1}{\sqrt{2}}\left|\Xi_c^{(\prime)+}D^{(*)-}\right\rangle-\frac{1}{\sqrt{2}}\left|\Xi_c^{(\prime)0}\bar{D}^{(*)0}\right\rangle$ \\
\hline
Systems&$\left|S,S_3\right\rangle$ & $\chi_{\rm{spin}}$ \\
\hline
\multirow{2}{*}{$\Xi_c^{(\prime)}\bar{D}$}&$\left|\frac{1}{2}, \frac{1}{2}\right\rangle$ & $\left|\frac{1}{2},\frac{1}{2}\right\rangle\left|0,0\right\rangle$ \\
&$\left|\frac{1}{2},-\frac{1}{2}\right\rangle$ & $\left|\frac{1}{2},-\frac{1}{2}\right\rangle\left|0,0\right\rangle$ \\
\multirow{6}{*}{$\Xi_c^{(\prime)}\bar{D}^{*}$}&$\left|\frac{1}{2}, \frac{1}{2}\right\rangle$ &$\frac{1}{\sqrt{3}}\left|\frac{1}{2},\frac{1}{2}\right\rangle\left|1,0\right\rangle-\sqrt{\frac{2}{3}}\left|\frac{1}{2},-\frac{1}{2}\right\rangle\left|1,1\right\rangle$ \\
&$\left|\frac{1}{2},-\frac{1}{2}\right\rangle$ & $\sqrt{\frac{2}{3}}\left|\frac{1}{2},\frac{1}{2}\right\rangle\left|1,-1\right\rangle-\frac{1}{\sqrt{3}}\left|\frac{1}{2},-\frac{1}{2}\right\rangle\left|1,0\right\rangle$ \\
&$\left|\frac{3}{2}, \frac{3}{2}\right\rangle$ & $\left|\frac{1}{2},\frac{1}{2}\right\rangle\left|1,1\right\rangle$ \\
&$\left|\frac{3}{2}, \frac{1}{2}\right\rangle$ & $\sqrt{\frac{2}{3}}\left|\frac{1}{2},\frac{1}{2}\right\rangle\left|1,0\right\rangle+\frac{1}{\sqrt{3}}\left|\frac{1}{2},-\frac{1}{2}\right\rangle\left|1,1\right\rangle$ \\
&$\left|\frac{3}{2},-\frac{1}{2}\right\rangle$ & $\frac{1}{\sqrt{3}}\left|\frac{1}{2},\frac{1}{2}\right\rangle\left|1,-1\right\rangle+\sqrt{\frac{2}{3}}\left|\frac{1}{2},-\frac{1}{2}\right\rangle\left|1,0\right\rangle$ \\
&$\left|\frac{3}{2},-\frac{3}{2}\right\rangle$ & $\left|\frac{1}{2},-\frac{1}{2}\right\rangle\left|1,-1\right\rangle$ \\
\bottomrule[1.0pt]
\bottomrule[1.0pt]
\end{tabular}
\end{table}

Similar to the magnetic moments and the transition magnetic moments of $\Xi_c^{(\prime)}$ and $\bar D^{(*)}$, the magnetic moments and the transition magnetic moments of the $S$-wave isoscalar $\Xi_c^{(\prime)}\bar D^{(*)}$ molecular states can be obtained by calculating the matrix elements $\mu_{H_0}=\left\langle{H_0} \left|\hat{\mu}_z \right| {H_0}\right\rangle$ and $\mu_{H_1 \to H_2}=\left\langle{H_2} \left|\hat{\mu}_z \right| {H_1}\right\rangle$, respectively. In Table \ref{Pwavefunctions}, we collect the flavor wave functions $\chi_{\rm{flavor}}$ and the spin wave functions $\chi_{\rm{spin}}$ for these discussed $S$-wave isoscalar $\Xi_c^{(\prime)}\bar D^{(*)}$ systems \cite{Wang:2022mxy}. Now, we take $\mu_{\Xi_{c} \bar D|^{2} S_{1/2}\rangle}$ and $\mu_{\Xi_{c}^{\prime} \bar D|^{2} S_{1 / 2}\rangle \to \Xi_{c} \bar D|^{2} S_{1 / 2}\rangle}$ as examples to illustrate the procedure of deducing the magnetic moments and the transition magnetic moments of the $S$-wave isoscalar $\Xi_c^{(\prime)}\bar D^{(*)}$ molecular states
\begin{eqnarray}
  \mu_{\Xi_{c} \bar D|^{2} S_{1/2}\rangle}&=&\left\langle \chi_{\Xi_{c} \bar D|^{2} S_{1/2}\rangle} \left|\hat{\mu}_z\right| \chi_{\Xi_{c} \bar D|^{2} S_{1/2}\rangle}\right\rangle\nonumber\\
  &=&\frac{1}{2} \mu_{\Xi_c^{+}}+\frac{1}{2} \mu_{\Xi_c^0},\label{SPmagneticmoment}\\
  \mu_{\Xi_{c}^{\prime} \bar D|^{2} S_{1 / 2}\rangle \to \Xi_{c} \bar D|^{2} S_{1 / 2}\rangle}
  &=&\left\langle \chi_{\Xi_{c} \bar D|^{2} S_{1 / 2}\rangle} \left|\hat{\mu}_z\right| \chi_{\Xi_{c}^{\prime} \bar D|^{2} S_{1 / 2}\rangle}\right\rangle \nonumber\\
  &=&\frac{1}{2} \mu_{\Xi_c^{\prime+} \to \Xi_c^{+}}+\frac{1}{2} \mu_{\Xi_c^{\prime0} \to \Xi_c^{0}},
\end{eqnarray}
respectively. Thus, we can obtain the magnetic moment of the $\Xi_{c} \bar D|^{2} S_{1/2}\rangle$ state and the transition magnetic moment of the ${\Xi_{c}^{\prime} \bar D|^{2} S_{1 / 2}\rangle \to \Xi_{c} \bar D|^{2} S_{1 / 2}\rangle} \gamma$ process.

After that, we further discuss the magnetic moments and the transition magnetic moments of the isoscalar $\Xi_c^{(\prime)}\bar D^{(*)}$ molecular states after considering the $S$-$D$ wave mixing effect. Nevertheless, the isoscalar $\Xi_{c}^{(\prime)}\bar D$ states with $J^P=1/2^-$ only exist the $S$-wave component, i.e., $|^{2} S_{1 / 2}\rangle$ channel, and the probabilities for the $D$-wave components are zero for the isoscalar $\Xi_{c}\bar D^*$ states with $J^P=1/2^-$ and $J^P=3/2^-$ \cite{Wang:2022mxy}. Thus for the $S$-$D$ wave mixing effects we consider the following allowed channels \cite{Wang:2022mxy}, i.e.,
\begin{eqnarray*}
&&\Xi_c^{\prime}\bar D^{*}[J^P=1/2^-]:|^{2} S_{1/2}\rangle,\,|^{4} D_{1/2}\rangle,\nonumber\\
&&\Xi_c^{\prime}\bar D^{*}[J^P=3/2^-]:|^{4} S_{3/2}\rangle,\,|^{2} D_{3/2}\rangle,\,|^{4} D_{3/2}\rangle.
\end{eqnarray*}
Here, the notation $|^{2S+1} L_{J}\rangle$ is used. Additionally, $S$, $L$, and $J$ denote the spin, orbit angular momentum, and total angular momentum quantum numbers for the discussed system, respectively.

When considering the contribution of the $S$-$D$ wave mixing effect, the hadronic magnetic moment can be written as $\sum_{i}\mu_{i}\langle R_{i}|R_{i}\rangle+\sum_{i\neq j} \mu_{i \to j}\langle R_{j}|R_{i}\rangle$, where $R_{i}$ denotes the space wave function of the corresponding $i$-th channel, which can be  obtained by the calculation of the mass spectrum. In the following, we take the $S$-wave isoscalar $\Xi_c^{\prime}\bar D^{*}$ state with $J^P=1/2^-$ as an example to illustrate how to obtain the hadronic magnetic moment after considering the $S$-$D$ wave mixing effect. After performing the $S$-$D$ wave mixing analysis, the specific expression of the magnetic moment of the $S$-wave isoscalar $\Xi_c^{\prime}\bar D^{*}$ state with $J^P=1/2^-$ can be written as
\begin{eqnarray}
&&\mu_{\Xi_c^{\prime}\bar D^{*}|^{2} S_{1/2}\rangle}\langle R_{\Xi_c^{\prime}\bar D^{*}|^{2} S_{1/2}\rangle}|R_{\Xi_c^{\prime}\bar D^{*}|^{2} S_{1/2}\rangle}\rangle\nonumber\\
&+&\mu_{\Xi_c^{\prime}\bar D^{*}|^{4} D_{1/2}\rangle}\langle R_{\Xi_c^{\prime}\bar D^{*}|^{4} D_{1/2}\rangle}|R_{\Xi_c^{\prime}\bar D^{*}|^{4} D_{1/2}\rangle}\rangle.
\end{eqnarray}
In the above expression, the $\mu_{\Xi_c^{\prime}\bar D^{*}|^{2} S_{1/2}\rangle}$ has been discussed in Eq. (\ref{SPmagneticmoment}), and the involved components $\langle R_{\Xi_c^{\prime}\bar D^{*}|^{2} S_{1/2}\rangle}|R_{\Xi_c^{\prime}\bar D^{*}|^{2} S_{1/2}\rangle}\rangle$ and $\langle R_{\Xi_c^{\prime}\bar D^{*}|^{4} D_{1/2}\rangle}|R_{\Xi_c^{\prime}\bar D^{*}|^{4} D_{1/2}\rangle}\rangle$ can be obtained by solving the Schr\"odinger equation for the mass spectrum of the isoscalar $\Xi_c^{\prime}\bar D^{*}$ state with $J^P=1/2^-$ after considering the $S$-$D$ wave mixing effect \cite{Wang:2022mxy}.

In the following, we illustrate the method of obtaining the magnetic moment of the $\Xi_c^{\prime}\bar D^{*}|^{4} D_{1/2}\rangle$ channel. For these investigated hadronic states, their spin-orbit wave function can be constructed as
\begin{eqnarray}
\left|{ }^{2 S+1} L_{J}\right\rangle=\sum_{m_{S}, m_{L}} C_{S m_{S}, L m_{L}}^{J, M} \chi_{S, m_{S}}Y_{L, m_{L}}.
\end{eqnarray}
Once expanding the spin-orbit wave function $\left|{ }^{4} D_{1 / 2}\right\rangle$ for the $\Xi_c^{\prime}\bar D^{*}$ system, we can obtain
\begin{eqnarray}
\left|{ }^{4} D_{1/2}\right\rangle &=&\frac{1}{\sqrt{10}}\chi_{{3}/{2},\,{3}/{2}} Y_{2,\,-1}-\frac{1}{\sqrt{5}}\chi_{{3}/{2},\,{1}/{2}} Y_{2,\,0}\nonumber\\
  &&+\sqrt{\frac{3}{10}}\chi_{{3}/{2},\,-{1}/{2}} Y_{2,\,1}-\sqrt{\frac{2}{5}}\chi_{{3}/{2},\,-{3}/{2}}Y_{2,\,2}.\nonumber\\
\end{eqnarray}
With the above preparation, the magnetic moment of the $\Xi_c^{\prime}\bar D^{*}|^{4} D_{1/2}\rangle$ channel can be deduced as
\begin{eqnarray}
&&\frac{1}{10}\left(\frac{1}{2} \mu_{\Xi_c^{\prime+}}+\frac{1}{2}\mu_{D^{*-}}+\frac{1}{2} \mu_{\Xi_c^{\prime0}}+\frac{1}{2}\mu_{\bar D^{*0}}-\mu_{\Xi_c^{\prime}\bar D^{*}}^L\right)\nonumber\\
&&+\frac{1}{5}\left(\frac{1}{6} \mu_{\Xi_c^{\prime+}}+\frac{1}{6}\mu_{D^{*-}}+\frac{1}{6} \mu_{\Xi_c^{\prime0}}+\frac{1}{6}\mu_{\bar D^{*0}}\right)\nonumber\\
&&+\frac{3}{10}\left(-\frac{1}{6} \mu_{\Xi_c^{\prime+}}-\frac{1}{6}\mu_{D^{*-}}-\frac{1}{6} \mu_{\Xi_c^{\prime0}}-\frac{1}{6}\mu_{\bar D^{*0}}+\mu_{\Xi_c^{\prime}\bar D^{*}}^L\right)\nonumber\\
&&+\frac{2}{5}\left(-\frac{1}{2} \mu_{\Xi_c^{\prime+}}-\frac{1}{2}\mu_{D^{*-}}-\frac{1}{2} \mu_{\Xi_c^{\prime0}}-\frac{1}{2}\mu_{\bar D^{*0}}+2\mu_{\Xi_c^{\prime}\bar D^{*}}^L\right)\nonumber\\
&=&-\frac{1}{6} \mu_{\Xi_c^{\prime+}}-\frac{1}{6}\mu_{D^{*-}}-\frac{1}{6} \mu_{\Xi_c^{\prime0}}-\frac{1}{6}\mu_{\bar D^{*0}}+\mu_{\Xi_c^{\prime}\bar D^{*}}^L.
\end{eqnarray}
Through the above calculation, we can get the magnetic moment of the isoscalar $\Xi_c^{\prime}\bar D^{*}$ state with $J^P=1/2^-$ after considering the $S$-$D$ wave mixing effect.

In addition, the general expression of the hadronic transition magnetic moment can be written as $\sum_{i\neq j} \mu_{i \rightarrow j}\left\langle R_{j}|R_{i}\right\rangle$ when considering the contribution of the $S$-$D$ wave mixing effect. Here, we need to mention that the calculation method of the transition magnetic moment is similar to that of the magnetic moment for the $D$-wave channel, except for the different wave functions of the initial and final states.

The radiative decay widths between the isoscalar $\Xi_c^{(\prime)}\bar D^{(*)}$ molecular states are the important physical observable in experiment, which may provide the crucial information to probe their inner structures. The radiative decay width is closely related to the transition magnetic moment. For the $H_{1} \to H_{2}\gamma$ process, the photon momentum can be written as \cite{Franklin:1981rc,Dey:1994qi,Hazra:2021lpa,Simonis:2018rld}
\begin{equation}
E_{\gamma}=\frac{M_1^2-M_2^2}{2M_1},\label{width1}
\end{equation}
where $H_{1}$ and $H_{2}$ stand for these discussed hidden-charm molecular pentaquarks with strangeness, while $M_1$ and $M_2$ are the masses of the hadrons $H_{1}$ and $ H_{2}$, respectively. The relation between the radiative decay width $\Gamma\left(H_{1} \to H_{2}\gamma\right)$ and the transition magnetic moment $\mu_{H_{1} \to H_{2}}$ can be given by \cite{Franklin:1981rc,Dey:1994qi,Hazra:2021lpa,Simonis:2018rld}
\begin{equation}
  \Gamma\left(H_{1} \to H_{2}\gamma\right)=\alpha_{\rm {EM}}\frac{E_{\gamma}^{3}}{M_{P}^{2}} \frac{2}{2J+1}\frac{\mu_{H_{1} \to H_{2}}^{2}}{\mu_N^{2}}. \label{width2}
\end{equation}
In the above expression, $\alpha_{\rm {EM}} \approx {1}/{137}$ is the fine structure constant, $M_P$ is the proton mass, and $J$ is the total angular momentum of the initial hidden-charm molecular pentaquarks with strangeness.

\section{Numerical analysis}\label{sec3}

In this section, we present the numerical results and discussions of the magnetic moments, the transition magnetic moments, and the radiative decay widths of the isoscalar $\Xi_c^{(\prime)}\bar D^{(*)}$ molecular states by performing the single channel analysis and the $S$-$D$ wave mixing analysis, respectively. In Table~\ref{Parameters}, the masses of these involved hadrons \cite{Workman:2022ynf} are collected.
\renewcommand\tabcolsep{0.31cm}
\renewcommand{\arraystretch}{1.50}
\begin{table}[!htbp]
\caption{The summary of the masses of these involved hadrons \cite{Workman:2022ynf}.}\label{Parameters}
\begin{tabular}{cc|cc}\toprule[1pt]\toprule[1pt]
Baryons & Masses\,(MeV) & Mesons & Masses\,(MeV)\\\hline
$\Xi_c^{+}$&2467.71          &$D^{-}$&1869.66\\
$\Xi_c^{0}$&2470.44          &$\bar D^{0}$&1864.84\\
$\Xi_c^{\prime+}$&2578.20    &$D^{*-}$&2010.26\\
$\Xi_c^{\prime0}$&2578.70    &$\bar D^{*0}$&2006.85\\
\bottomrule[1pt]\bottomrule[1pt]
\end{tabular}
\end{table}

\renewcommand\tabcolsep{0.08cm}
\renewcommand{\arraystretch}{1.50}
\begin{table}[!htbp]
  \caption{The magnetic moments of the $S$-wave isoscalar $\Xi_c^{(\prime)}\bar D^{(*)}$-type hidden-charm molecular pentaquarks with strangeness obtained by performing the single channel analysis. Here, the magnetic moment is in unit of the nuclear magnetic moment $\mu_N$. }
  \label{PM}
\begin{tabular}{l|c|cc}
\toprule[1.0pt]
\toprule[1.0pt]
States  & $I(J^P)$ & Expressions & Results \\\hline
$\Xi_c\bar{D}$ & $0(\frac{1}{2}^-)$ & $\frac{1}{2}(\mu_{\Xi_c^{+}}+\mu_{\Xi_c^{0}})$ & $0.372$ \\
$\Xi_c^{\prime}\bar{D}$ & $0(\frac{1}{2}^-)$ & $\frac{1}{2}(\mu_{\Xi_c^{\prime+}}+\mu_{\Xi_c^{\prime0}})$ & $-0.277$ \\
\multirow{2}{*}{$\Xi_c\bar{D}^{*}$} & $0(\frac{1}{2}^-)$ & $-\frac{1}{6}(\mu_{\Xi_c^{+}}+\mu_{\Xi_c^{0}})+\frac{1}{3}(\mu_{D^{*-}}+\mu_{\bar{D}^{*0}})$ & $-0.062$\\
                   & $0(\frac{3}{2}^-)$ & $\frac{1}{2}(\mu_{\Xi_c^{+}}+\mu_{D^{*-}}+\mu_{\Xi_c^{0}}+\mu_{\bar{D}^{*0}})$ & $0.465$\\
\multirow{2}{*}{$\Xi_c^{\prime}\bar{D}^{*}$} &$0(\frac{1}{2}^-)$ & $-\frac{1}{6}(\mu_{\Xi_c^{\prime+}}+\mu_{\Xi_c^{\prime0}})+\frac{1}{3}(\mu_{D^{*-}}+\mu_{\bar{D}^{*0}})$& $0.154$\\
                            &$0(\frac{3}{2}^-)$ & $\frac{1}{2}(\mu_{\Xi_c^{\prime+}}+\mu_{D^{*-}}+\mu_{\Xi_c^{\prime0}}+\mu_{\bar{D}^{*0}})$ & $-0.184$\\
\bottomrule[1.0pt]
\bottomrule[1.0pt]
\end{tabular}
\end{table}

\begin{figure}[!htbp]
  \includegraphics[width=0.48\textwidth]{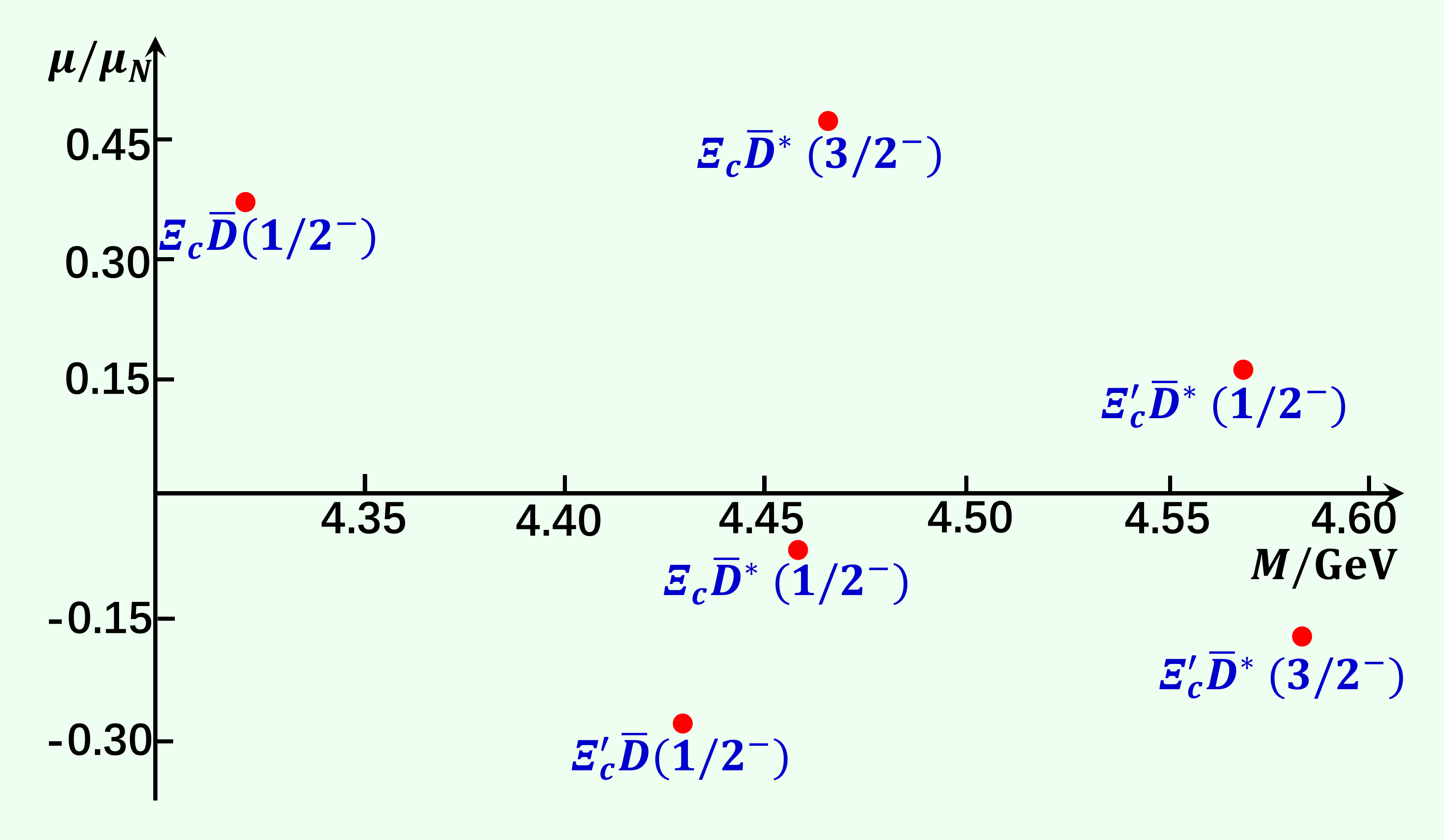}
  \caption{The masses and magnetic moments of the $S$-wave isoscalar $\Xi_c^{(\prime)}\bar D^{(*)}$ molecular states. Here, we estimate the mass positions of these hidden-charm molecular pentaquarks with strangeness based on the results from Refs. \cite{Wang:2022mxy,Wang:2019nvm,Chen:2021spf}.}\label{MM}
\end{figure}

We first give the expressions and numerical results for the magnetic moments of the $S$-wave isoscalar hidden-charm $\Xi_c^{(\prime)}\bar D^{(*)}$ molecular states with strangeness in Table \ref{PM}. Here, we mainly want to answer whether or not the properties of the magnetic moments can be applied to distinguish the $S$-wave isoscalar $\Xi_c^{(\prime)}\bar D^{*}$ molecular states with $J^P=1/2^-$ and $J^P=3/2^-$. To be more intuitively, we present their masses \cite{Wang:2022mxy,Wang:2019nvm,Chen:2021spf} and magnetic moments in Fig. \ref{MM}.

Based on the obtained numerical results, we summarize the following points:
\begin{itemize}
  \item Since the magnetic moment of $\bar D$ is zero, those of the $S$-wave isoscalar $\Xi_c^{(\prime)}\bar D$ molecules only contain the contribution from $\Xi_c^{(\prime)}$.
  \item The magnetic moment of the $S$-wave isoscalar $\Xi_c\bar D^{*}$ molecule with $J^P=1/2^-$ is $-0.062~\mu_{N}$, while those with $J^P=3/2^-$ is $0.465~\mu_{N}$. They are much different, and this can help us to distinguish whether the molecular states are with $J^P=1/2^-$ and $J^P=3/2^-$, which deserves to attract the attentions in the future experiments.
  \item Similar to the $\Xi_c\bar D^{*}$ case, the magnetic moment of the $S$-wave isoscalar $\Xi_c^{\prime}\bar D^{*}$ molecule with $J^P=1/2^-$ is obviously different from that with $J^P=3/2^-$. If the magnetic moment sign is measured as positive in the future experiments, the spin-parity quantum number of the $S$-wave isoscalar $\Xi_c^{\prime}\bar D^{*}$ molecular state would be $J^P=1/2^-$ rather than $J^P=3/2^-$.
  \item The $S$-wave isoscalar $\Xi_c\bar D^{(*)}$ and $\Xi_c^{\prime}\bar D^{(*)}$ states with the same spin-parity quantum numbers have same spin wave functions (see Table \ref{Pwavefunctions} for more details), but their magnetic moments are different, which is because the magnetic moment of $\Xi_c$ is significantly different from that of $\Xi_c^{\prime}$. Therefore, studying the magnetic moments can provide useful hints to probe the inner structures of these isoscalar $\Xi_c^{(\prime)}\bar D^{(*)}$ molecular states.
\end{itemize}

Through the above analysis, we can conclude that the discussion of the magnetic moments can provide crucial information to understand their inner structures and distinguish the $S$-wave isoscalar $\Xi_c^{(\prime)}\bar D^{*}$ molecular states with $J^P=1/2^-$ and $J^P=3/2^-$. To some extent, this fact shows the importance of exploring the hadronic magnetic moments, and we wish that more experimental and theoretical colleagues pay more attention to discussing the magnetic moment properties of the $S$-wave isoscalar $\Xi_c^{(\prime)}\bar D^{(*)}$ molecular states in the near future.

And then, we discuss the transition magnetic moments and the corresponding radiative decay widths between the $S$-wave isoscalar $\Xi_c^{(\prime)}\bar D^{(*)}$ molecular states, which include:
\begin{eqnarray}
&&\Xi_c^{\prime}\bar{D}|{1}/{2}^-\rangle \to \Xi_c\bar{D}|{1}/{2}^-\rangle\gamma, \nonumber\\
&&\Xi_c\bar{D}^{*}|{1}/{2}^-\rangle \to \Xi_c\bar{D}|{1}/{2}^-\rangle\gamma, \nonumber\\
&&
\Xi_c\bar{D}^{*}|{3}/{2}^-\rangle \to \Xi_c\bar{D}|{1}/{2}^-\rangle\gamma, \nonumber\\
&&
\Xi_c^{\prime}\bar{D}^{*}|{1}/{2}^-\rangle \to \Xi_c\bar{D}|{1}/{2}^-\rangle\gamma, \nonumber\\
&&
\Xi_c^{\prime}\bar{D}^{*}|{3}/{2}^-\rangle \to \Xi_c\bar{D}|{1}/{2}^-\rangle\gamma, \nonumber\\
&&
\Xi_c\bar{D}^{*}|{1}/{2}^-\rangle \to \Xi_c^{\prime}\bar{D}|{1}/{2}^-\rangle\gamma, \nonumber\\
&&
\Xi_c\bar{D}^{*}|{3}/{2}^-\rangle \to \Xi_c^{\prime}\bar{D}|{1}/{2}^-\rangle \gamma, \nonumber\\
&&
\Xi_c^{\prime}\bar{D}^{*}|{1}/{2}^-\rangle \to \Xi_c^{\prime}\bar{D}|{1}/{2}^-\rangle\gamma, \nonumber\\
&&
\Xi_c^{\prime}\bar{D}^{*}|{3}/{2}^-\rangle \to \Xi_c^{\prime}\bar{D}|{1}/{2}^-\rangle\gamma, \nonumber\\
&&
\Xi_c\bar{D}^{*}|{3}/{2}^-\rangle \to \Xi_c\bar{D}^{*}|{1}/{2}^-\rangle \gamma, \nonumber\\
&&
\Xi_c^{\prime}\bar{D}^{*}|{1}/{2}^-\rangle \to \Xi_c\bar{D}^{*}|{1}/{2}^-\rangle\gamma, \nonumber\\
&&
\Xi_c^{\prime}\bar{D}^{*}|{3}/{2}^-\rangle \to \Xi_c\bar{D}^{*}|{1}/{2}^-\rangle\gamma, \nonumber\\
&&
\Xi_c^{\prime}\bar{D}^{*}|{1}/{2}^-\rangle \to \Xi_c\bar{D}^{*}|{3}/{2}^-\rangle\gamma, \nonumber\\
&&
\Xi_c^{\prime}\bar{D}^{*}|{3}/{2}^-\rangle \to \Xi_c\bar{D}^{*}|{3}/{2}^-\rangle\gamma, \nonumber\\
&&
\Xi_c^{\prime}\bar{D}^{*}|{3}/{2}^-\rangle \to \Xi_c^{\prime}\bar{D}^{*}|{1}/{2}^-\rangle\gamma.\nonumber
\end{eqnarray}
Similar to the magnetic moments of the $S$-wave isoscalar $\Xi_c^{(\prime)}\bar D^{*}$ molecular states, we expect that the study of the transition magnetic moments and the radiative decay behaviors can provide important hints to disclose their inner structures and distinguish the $S$-wave isoscalar $\Xi_c^{(\prime)}\bar D^{*}$ molecules with $J^P=1/2^-$ and $J^P=3/2^-$.

In Table \ref{PTM}, we present  the expressions and numerical results of the transition magnetic moments between the $S$-wave isoscalar $\Xi_c^{(\prime)}\bar D^{(*)}$-type hidden-charm molecular pentaquarks with strangeness. Among them, the largest and smallest transition magnetic moments correspond to the $\Xi_c\bar{D}^{*}|{3}/{2}^-\rangle \to \Xi_c\bar{D}^{*}|{1}/{2}^-\rangle\gamma$ and the $\Xi_c^{(\prime)}\bar{D}^{*}|{3}/{2}^-\rangle \to \Xi_c^{(\prime)}\bar{D}|{1}/{2}^-\rangle\gamma$ processes, respectively.
\renewcommand\tabcolsep{1.30cm}
\renewcommand{\arraystretch}{1.50}
\begin{table*}[!htbp]
  \caption{The transition magnetic moments between the $S$-wave isoscalar $\Xi_c^{(\prime)}\bar D^{(*)}$-type hidden-charm molecular pentaquarks with strangeness obtained by performing the single channel analysis. Here, the magnetic moment is in unit of the nuclear magnetic moment $\mu_N$.}
  \label{PTM}
\begin{tabular}{c|cc}
\toprule[1.0pt]
\toprule[1.0pt]
Decay modes & Expressions & Results \\
\hline
$\Xi_c^{\prime}\bar{D}|\frac{1}{2}^-\rangle \to \Xi_c\bar{D}|\frac{1}{2}^-\rangle\gamma$ & $\frac{1}{2}(\mu_{\Xi_c^{\prime+} \to \Xi_c^{+}}+\mu_{\Xi_c^{\prime0} \to \Xi_c^{0}})$ & $-0.670$  \\
$\Xi_c\bar{D}^{*}|\frac{1}{2}^-\rangle \to \Xi_c\bar{D}|\frac{1}{2}^-\rangle\gamma$ & $\frac{1}{2\sqrt{3}}(\mu_{D^{*-} \to D^{-}}+\mu_{\bar{D}^{*0} \to \bar{D}^{0}})$ & $-0.484$ \\
$\Xi_c\bar{D}^{*}|\frac{3}{2}^-\rangle \to \Xi_c\bar{D}|\frac{1}{2}^-\rangle\gamma$ & $\frac{1}{\sqrt{6}}(\mu_{D^{*-} \to D^{-}}+\mu_{\bar{D}^{*0} \to \bar{D}^{0}})$ & $-0.684$ \\
$\Xi_c^{\prime}\bar{D}^{*}|\frac{1}{2}^-\rangle \to \Xi_c^{\prime}\bar{D}|\frac{1}{2}^-\rangle\gamma$ & $\frac{1}{2\sqrt{3}}(\mu_{D^{*-} \to D^{-}}+\mu_{\bar{D}^{*0} \to \bar{D}^{0}})$ & $-0.484$ \\
$\Xi_c^{\prime}\bar{D}^{*}|\frac{3}{2}^-\rangle \to \Xi_c^{\prime}\bar{D}|\frac{1}{2}^-\rangle\gamma$ & $\frac{1}{\sqrt{6}}(\mu_{D^{*-} \to D^{-}}+\mu_{\bar{D}^{*0} \to \bar{D}^{0}})$ & $-0.684$ \\
$\Xi_c\bar{D}^{*}|\frac{3}{2}^-\rangle \to \Xi_c\bar{D}^{*}|\frac{1}{2}^-\rangle\gamma$ & $\frac{\sqrt{2}}{3}(\mu_{\Xi_c^{+}}+\mu_{\Xi_c^{0}})-\frac{1}{3\sqrt{2}}(\mu_{D^{*-}}+\mu_{\bar{D}^{*0}})$ & $0.307$ \\
$\Xi_c^{\prime}\bar{D}^{*}|\frac{1}{2}^-\rangle \to \Xi_c\bar{D}^{*}|\frac{1}{2}^-\rangle\gamma$ & $-\frac{1}{6}(\mu_{\Xi_c^{\prime+} \to \Xi_c^{+}}+\mu_{\Xi_c^{\prime0} \to \Xi_c^{0}})$ & $0.223$  \\
$\Xi_c^{\prime}\bar{D}^{*}|\frac{3}{2}^-\rangle \to \Xi_c\bar{D}^{*}|\frac{1}{2}^-\rangle\gamma$ & $\frac{\sqrt{2}}{3}(\mu_{\Xi_c^{\prime+} \to \Xi_c^{+}}+\mu_{\Xi_c^{\prime0} \to \Xi_c^{0}})$ & $-0.632$  \\
$\Xi_c^{\prime}\bar{D}^{*}|\frac{1}{2}^-\rangle \to \Xi_c\bar{D}^{*}|\frac{3}{2}^-\rangle\gamma$ & $\frac{\sqrt{2}}{3}(\mu_{\Xi_c^{\prime+} \to \Xi_c^{+}}+\mu_{\Xi_c^{\prime0} \to \Xi_c^{0}})$ & $-0.632$  \\
$\Xi_c^{\prime}\bar{D}^{*}|\frac{3}{2}^-\rangle \to \Xi_c\bar{D}^{*}|\frac{3}{2}^-\rangle\gamma$ & $\frac{1}{2}(\mu_{\Xi_c^{\prime+} \to \Xi_c^{+}}+\mu_{\Xi_c^{\prime0} \to \Xi_c^{0}})$ & $-0.670$  \\
$\Xi_c^{\prime}\bar{D}^{*}|\frac{3}{2}^-\rangle \to \Xi_c^{\prime}\bar{D}^{*}|\frac{1}{2}^-\rangle\gamma$ & $\frac{\sqrt{2}}{3}(\mu_{\Xi_c^{\prime+}}+\mu_{\Xi_c^{\prime0}})-\frac{1}{3\sqrt{2}}(\mu_{D^{*-}}+\mu_{\bar{D}^{*0}})$ & $-0.305$ \\
\bottomrule[1.0pt]
\bottomrule[1.0pt]
\end{tabular}
\end{table*}

By performing numerical calculation, we can find that the transition magnetic moments between the $S$-wave isoscalar $\Xi_c^{(\prime)}\bar D^{(*)}$ molecular states exist following important relations, i.e.,
\begin{eqnarray}
\frac{\mu_{\Xi_c\bar{D}^{*}|\frac{1}{2}^-\rangle \to \Xi_c\bar{D}|\frac{1}{2}^-\rangle}}{\mu_{\Xi_c\bar{D}^{*}|\frac{3}{2}^-\rangle \to \Xi_c\bar{D}|\frac{1}{2}^-\rangle}}&=&\frac{1}{\sqrt{2}},\nonumber\\
\frac{\mu_{\Xi_c^{\prime}\bar{D}^{*}|\frac{1}{2}^-\rangle \to \Xi_c^{\prime}\bar{D}|\frac{1}{2}^-\rangle}}{\mu_{\Xi_c^{\prime}\bar{D}^{*}|\frac{3}{2}^-\rangle \to \Xi_c^{\prime}\bar{D}|\frac{1}{2}^-\rangle}}&=&\frac{1}{\sqrt{2}},\nonumber\\
\frac{\mu_{\Xi_c^{\prime}\bar{D}^{*}|\frac{1}{2}^-\rangle \to \Xi_c\bar{D}^{*}|\frac{1}{2}^-\rangle}}{\mu_{\Xi_c^{\prime}\bar{D}^{*}|\frac{3}{2}^-\rangle \to \Xi_c\bar{D}^{*}|\frac{1}{2}^-\rangle}}&=&-\frac{1}{2\sqrt{2}},\nonumber\\
\frac{\mu_{\Xi_c^{\prime}\bar{D}^{*}|\frac{1}{2}^-\rangle \to \Xi_c\bar{D}^{*}|\frac{3}{2}^-\rangle}}{\mu_{\Xi_c^{\prime}\bar{D}^{*}|\frac{3}{2}^-\rangle \to \Xi_c\bar{D}^{*}|\frac{3}{2}^-\rangle}}&=&\frac{2\sqrt{2}}{3},\label{TMrelation}
\end{eqnarray}
which are determined by the flavor-spin wave functions of the initial and final states. From Eq. (\ref{TMrelation}), we can see that there are differences between the transition magnetic moments of the $S$-wave isoscalar $\Xi_c^{(\prime)}\bar D^{*}$ molecules with $J^P=1/2^-$ and $J^P=3/2^-$, and these important relations provide us the critical information to identify the spin-parity quantum numbers of the $S$-wave isoscalar $\Xi_c^{(\prime)}\bar D^{*}$ molecular states.

More interestingly, the transition magnetic moments between the $S$-wave isoscalar $\Xi_c^{(\prime)}\bar D^{(*)}$ molecular states exist several equality relations, such as
\begin{eqnarray}
\frac{\mu_{\Xi_c^{\prime}\bar{D}|\frac{1}{2}^-\rangle \to \Xi_c\bar{D}|\frac{1}{2}^-\rangle}}{\mu_{\Xi_c^{\prime}\bar{D}^{*}|\frac{3}{2}^-\rangle \to \Xi_c\bar{D}^{*}|\frac{3}{2}^-\rangle}}&=&1,\nonumber\\
\frac{\mu_{\Xi_c\bar{D}^{*}|\frac{1}{2}^-\rangle \to \Xi_c\bar{D}|\frac{1}{2}^-\rangle}}{\mu_{\Xi_c^{\prime}\bar{D}^{*}|\frac{1}{2}^-\rangle \to \Xi_c^{\prime}\bar{D}|\frac{1}{2}^-\rangle}}&=&1,\nonumber\\
\frac{\mu_{\Xi_c\bar{D}^{*}|\frac{3}{2}^-\rangle \to \Xi_c\bar{D}|\frac{1}{2}^-\rangle}}{\mu_{\Xi_c^{\prime}\bar{D}^{*}|\frac{3}{2}^-\rangle \to \Xi_c^{\prime}\bar{D}|\frac{1}{2}^-\rangle}}&=&1,\nonumber\\
\frac{\mu_{\Xi_c^{\prime}\bar{D}^{*}|\frac{3}{2}^-\rangle \to \Xi_c\bar{D}^{*}|\frac{1}{2}^-\rangle}}{\mu_{\Xi_c^{\prime}\bar{D}^{*}|\frac{1}{2}^-\rangle \to \Xi_c\bar{D}^{*}|\frac{3}{2}^-\rangle}}&=&1.\label{equalityrelations}
\end{eqnarray}
Therefore, we hope that such qualitative relations of Eqs. (\ref{TMrelation})-(\ref{equalityrelations}) can be further tested in the future experiments and other theoretical approaches, which also can be regarded as important relations to check our theoretical predictions.

After getting the transition magnetic moments between the $S$-wave isoscalar $\Xi_c^{(\prime)}\bar D^{(*)}$ molecular states, we further discuss the radiative decay behaviors between the $S$-wave isoscalar $\Xi_c^{(\prime)}\bar D^{(*)}$ molecular states. As illustrated in Eqs. (\ref{width1})-(\ref{width2}), the radiative decay widths depend on the binding energies of the initial and final molecules. For simplicity, we adopt the same binding energies for the initial and final molecules and take $-1.0$ MeV, $-6.0$ MeV, and $-12.0$ MeV to discuss the radiative decay widths between the $S$-wave isoscalar $\Xi_c^{(\prime)}\bar D^{(*)}$ molecular states in the absence of experimental data, which is similar to present the bound properties of the $S$-wave isoscalar $\Xi_c^{(\prime)}\bar D^{(*)}$ states in Ref. \cite{Wang:2022mxy}. In Table \ref{PD},  we show the numerical results of the radiative decay widths between the $S$-wave isoscalar $\Xi_c^{(\prime)}\bar D^{(*)}$-type hidden-charm molecular pentaquarks with strangeness.
\renewcommand\tabcolsep{0.06cm}
\renewcommand{\arraystretch}{1.50}
\begin{table}[!htbp]
  \caption{The radiative decay widths between the $S$-wave isoscalar $\Xi_c^{(\prime)}\bar D^{(*)}$-type hidden-charm molecular pentaquarks with strangeness obtained by performing the single channel analysis. Here, the radiative decay width is in unit of keV.}
  \label{PD}
\begin{tabular}{c|ccc}
\toprule[1.0pt]
\toprule[1.0pt]
{Decay modes} & $-1.0$ MeV & $-6.0$ MeV & $-12.0$ MeV \\
\cline{1-4}
$\Xi_c^{\prime}\bar{D}|\frac{1}{2}^-\rangle \to \Xi_c\bar{D}|\frac{1}{2}^-\rangle\gamma$ & $4.710$ & $4.694$ & $4.675$  \\
$\Xi_c\bar{D}^{*}|\frac{1}{2}^-\rangle \to \Xi_c\bar{D}|\frac{1}{2}^-\rangle\gamma$ & $5.239$ & $5.221$ & $5.200$ \\
$\Xi_c\bar{D}^{*}|\frac{3}{2}^-\rangle \to \Xi_c\bar{D}|\frac{1}{2}^-\rangle\gamma$ & $5.239$ & $5.221$ & $5.200$ \\
$\Xi_c^{\prime}\bar{D}^{*}|\frac{1}{2}^-\rangle \to \Xi_c^{\prime}\bar{D}|\frac{1}{2}^-\rangle\gamma$ & $5.244$ & $5.227$ & $5.206$ \\
$\Xi_c^{\prime}\bar{D}^{*}|\frac{3}{2}^-\rangle \to \Xi_c^{\prime}\bar{D}|\frac{1}{2}^-\rangle\gamma$ & $5.244$ & $5.227$ & $5.206$ \\
$\Xi_c^{\prime}\bar{D}^{*}|\frac{1}{2}^-\rangle \to \Xi_c\bar{D}^{*}|\frac{1}{2}^-\rangle\gamma$ & $0.524$ & $0.522$ & $0.520$  \\
$\Xi_c^{\prime}\bar{D}^{*}|\frac{3}{2}^-\rangle \to \Xi_c\bar{D}^{*}|\frac{1}{2}^-\rangle\gamma$ & $2.096$ & $2.089$ & $2.080$  \\
$\Xi_c^{\prime}\bar{D}^{*}|\frac{1}{2}^-\rangle \to \Xi_c\bar{D}^{*}|\frac{3}{2}^-\rangle\gamma$ & $4.191$ & $4.177$ & $4.161$  \\
$\Xi_c^{\prime}\bar{D}^{*}|\frac{3}{2}^-\rangle \to \Xi_c\bar{D}^{*}|\frac{3}{2}^-\rangle\gamma$ & $2.357$ & $2.350$ & $2.340$  \\
\bottomrule[1.0pt]
\bottomrule[1.0pt]
\end{tabular}
\end{table}

For these obtained radiative decay widths between the $S$-wave isoscalar $\Xi_c^{(\prime)}\bar D^{(*)}$ molecules, we need to point out that the binding energies of the $S$-wave isoscalar $\Xi_c^{(\prime)}\bar D^{(*)}$ moleculur states are very small compared with their mass thresholds, which can explain why the binding energies of the $S$-wave isoscalar $\Xi_c^{(\prime)}\bar D^{(*)}$ molecules play a minor role for their radiative decay widths \cite{Zhou:2022gra}. Furthermore, most of the radiative decay widths between the $S$-wave isoscalar $\Xi_c^{(\prime)}\bar D^{(*)}$ molecular states are around 5.0 keV, while the decay width of the $\Xi_c^{\prime}\bar{D}^{*}|{1}/{2}^-\rangle \to \Xi_c\bar{D}^{*}|{1}/{2}^-\rangle\gamma$ process is less than 1.0 keV, which is because the transition magnetic moment is small in this process.

When taking same binding energies for the $S$-wave isoscalar $\Xi_c\bar D^{*}$ molecular states with $J^P=1/2^-$ and $J^P=3/2^-$, we find that there exist same radiative decay widths for the $\Xi_c\bar{D}^{*}|{1}/{2}^-\rangle \to \Xi_c\bar{D}|{1}/{2}^-\rangle\gamma$ and $\Xi_c\bar{D}^{*}|{3}/{2}^-\rangle \to \Xi_c\bar{D}|{1}/{2}^-\rangle\gamma$ processes, since both radiative decay processes have same $\mu_{H_1 \to H_2}^2/(2J+1)$. As stressed in Ref. \cite{Wang:2022mxy}, the binding energies of the $S$-wave isoscalar $\Xi_c\bar D^{*}$ molecular states with $J^P=1/2^-$ and $J^P=3/2^-$ can be different slightly, which may result in the small difference of the radiative decay widths for the $\Xi_c\bar{D}^{*}|{1}/{2}^-\rangle \to \Xi_c\bar{D}|{1}/{2}^-\rangle\gamma$ and $\Xi_c\bar{D}^{*}|{3}/{2}^-\rangle \to \Xi_c\bar{D}|{1}/{2}^-\rangle\gamma$ processes. Here, we need to mention that the $P_{cs}(4459)$ existing in the $J/\psi \Lambda$ invariant mass spectrum may be described by two peak structures, and the corresponding masses are around 4454.9 MeV and 4467.9 MeV, respectively \cite{LHCb:2020jpq}. If the masses of the $S$-wave isoscalar $\Xi_c\bar D^{*}$ molecular states with $J^P=1/2^-$ and $J^P=3/2^-$ are 4454.9 MeV and 4467.9 MeV, we can estimate
\begin{eqnarray*}
\Gamma(\Xi_c\bar{D}^{*}|{1}/{2}^-\rangle \to \Xi_c\bar{D}|{1}/{2}^-\rangle\gamma)&=&3.596~{\rm keV},\nonumber\\
\Gamma(\Xi_c\bar{D}^{*}|{3}/{2}^-\rangle \to \Xi_c\bar{D}|{1}/{2}^-\rangle\gamma)&=&4.813~{\rm keV},
\end{eqnarray*}
respectively. Meanwhile, it is interesting to note that the radiative decay behaviors of the $\Xi_c^{\prime}\bar{D}^{*}|{1}/{2}^-\rangle \to \Xi_c\bar{D}|{1}/{2}^-\rangle\gamma$ and $\Xi_c^{\prime}\bar{D}^{*}|{3}/{2}^-\rangle \to \Xi_c\bar{D}|{1}/{2}^-\rangle\gamma$ processes are similar to that of the $\Xi_c\bar{D}^{*}|{1}/{2}^-\rangle \to \Xi_c\bar{D}|{1}/{2}^-\rangle\gamma$ and $\Xi_c\bar{D}^{*}|{3}/{2}^-\rangle \to \Xi_c\bar{D}|{1}/{2}^-\rangle\gamma$ processes, which is due to the similarity between the $S$-wave isoscalar $\Xi_c\bar D^{(*)}$ states and the $S$-wave isoscalar $\Xi_c^{\prime}\bar D^{(*)}$ states.

\begin{figure}[!htbp]
  \includegraphics[width=0.45\textwidth]{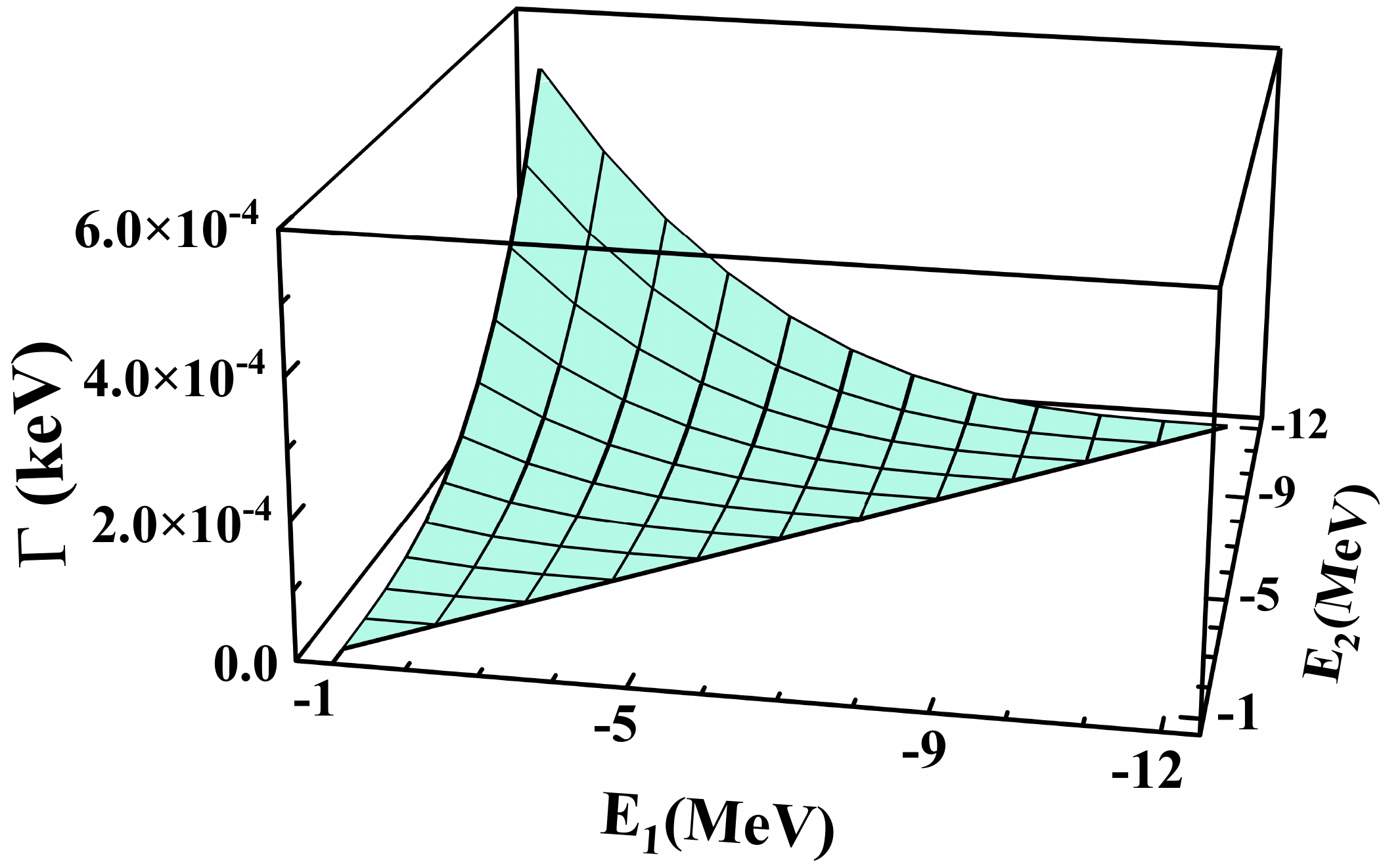}
  \caption{The dependence of the radiative decay width $\Gamma$ of the $\Xi_c\bar{D}^{*}|{3}/{2}^-\rangle \to \Xi_c\bar{D}^{*}|{1}/{2}^-\rangle\gamma$ process on the binding energy $E_1$ of the $\Xi_c\bar{D}^{*}$ molecule with $I(J^P)=0({1/2}^-)$ and the binding energy $E_2$ of the $\Xi_c\bar{D}^{*}$ molecule with $I(J^P)=0({3/2}^-)$. Here, the transition magnetic moment of the $\Xi_c\bar{D}^{*}|{3}/{2}^-\rangle \to \Xi_c\bar{D}^{*}|{1}/{2}^-\rangle\gamma$ process is taken as $0.307~\mu_N$.}\label{EEGammaXiDs}
\end{figure}

\begin{figure}[!htbp]
  \includegraphics[width=0.45\textwidth]{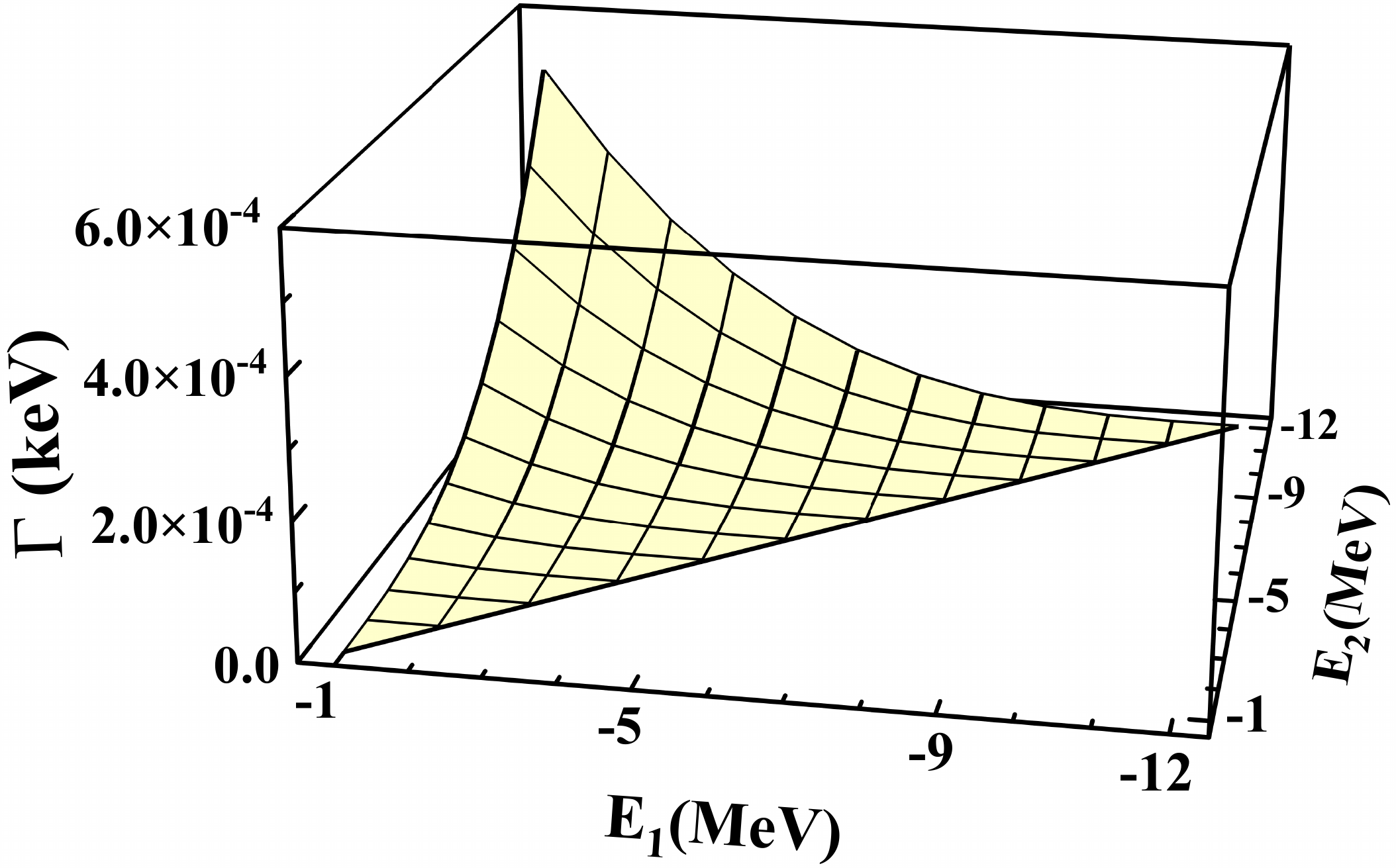}
  \caption{The dependence of the radiative decay width $\Gamma$ of the $\Xi_c^{\prime}\bar{D}^{*}|{3}/{2}^-\rangle \to \Xi_c^{\prime}\bar{D}^{*}|{1}/{2}^-\rangle\gamma$ process on the binding energy $E_1$ of the $\Xi_c^{\prime}\bar{D}^{*}$ molecule with $I(J^P)=0({1/2}^-)$ and the binding energy $E_2$ of the $\Xi_c^{\prime}\bar{D}^{*}$ molecule with $I(J^P)=0({3/2}^-)$. Here, the transition magnetic moment of the $\Xi_c^{\prime}\bar{D}^{*}|{3}/{2}^-\rangle \to \Xi_c^{\prime}\bar{D}^{*}|{1}/{2}^-\rangle\gamma$ process is taken as $-0.305~\mu_N$.}\label{EEGammaXipDs}
\end{figure}

In addition, when adopting same binding energies for the $S$-wave isoscalar $\Xi_c^{(\prime)}\bar D^{*}$ molecular states with $J^P=1/2^-$ and $J^P=3/2^-$, it is obvious that the radiative decay widths are zero for the $\Xi_c\bar{D}^{*}|{3}/{2}^-\rangle \to \Xi_c\bar{D}^{*}|{1}/{2}^-\rangle\gamma$ and $\Xi_c^{\prime}\bar{D}^{*}|{3}/{2}^-\rangle \to \Xi_c^{\prime}\bar{D}^{*}|{1}/{2}^-\rangle\gamma$ processes. In practice, the binding energies of the $S$-wave isoscalar $\Xi_c^{(\prime)}\bar D^{*}$ molecular states with $J^P=1/2^-$ and $J^P=3/2^-$ can exist difference \cite{Wang:2022mxy}. In Figs. \ref{EEGammaXiDs} and \ref{EEGammaXipDs}, we present the radiative decay widths of the $\Xi_c\bar{D}^{*}|{3}/{2}^-\rangle \to \Xi_c\bar{D}^{*}|{1}/{2}^-\rangle\gamma$ and $\Xi_c^{\prime}\bar{D}^{*}|{3}/{2}^-\rangle \to \Xi_c^{\prime}\bar{D}^{*}|{1}/{2}^-\rangle\gamma$ processes with different binding energies for the $S$-wave isoscalar $\Xi_c^{(\prime)}\bar D^{*}$ molecules with $J^P=1/2^-$ and $J^P=3/2^-$, respectively. According to our quantitative calculation, the radiative decay widths of the $\Xi_c\bar{D}^{*}|{3}/{2}^-\rangle \to \Xi_c\bar{D}^{*}|{1}/{2}^-\rangle\gamma$ and $\Xi_c^{\prime}\bar{D}^{*}|{3}/{2}^-\rangle \to \Xi_c^{\prime}\bar{D}^{*}|{1}/{2}^-\rangle\gamma$ processes are less than 0.001 keV. The main reason is that the hadronic molecule is a loosely bound state with the reasonable binding energy at most tens of MeV \cite{Chen:2016qju}, and the masses of the $S$-wave isoscalar $\Xi_c^{(\prime)}\bar D^{*}$ states with $J^P=1/2^-$ and $J^P=3/2^-$ are extremely close to each other in the hadronic molecular picture \cite{Wang:2022mxy}, which leads to the  suppression of the kinetic phase space for both radiative decay processes. Furthermore, if the masses of the $S$-wave isoscalar $\Xi_c\bar D^{*}$ molecular states with $J^P=1/2^-$ and $J^P=3/2^-$ are taken as 4454.9 MeV and 4467.9 MeV \cite{LHCb:2020jpq}, the radiative decay width of the $\Xi_c\bar{D}^{*}|{3}/{2}^-\rangle \to \Xi_c\bar{D}^{*}|{1}/{2}^-\rangle\gamma$ process is predicted to be around $0.0008~{\rm keV}$.

In the above discussion, we mainly focus on the magnetic moments, the transition magnetic moments, and the radiative decay behaviors of the $S$-wave isoscalar $\Xi_c^{(\prime)}\bar D^{(*)}$ molecular states by performing the single channel analysis, which provides valuable information to reflect their inner structures. Next, we further discuss them with considering the $S$-$D$ wave mixing effect. According to our previous results on the mass spectrum \cite{Wang:2022mxy}, we find that the $S$-wave channels have the dominant contribution with the probabilities over $95$ percent and play a major role for the isoscalar $\Xi_c^{\prime}\bar D^{*}$ molecular states with $J^P=1/2^-$ and $J^P=3/2^-$. Thus, we naturally conjecture that the $D$-wave components with small contributions do not obviously decorate the magnetic moments of the $S$-wave isoscalar $\Xi_c^{\prime}\bar D^{*}$ molecular states with $J^P=1/2^-$ and $J^P=3/2^-$ and the transition magnetic moment of the $\Xi_c^{\prime}\bar D^{*}|{3}/{2}^-\rangle  \to \Xi_c^{\prime}\bar D^{*}|{1}/{2}^-\rangle \gamma$ process. Indeed, our following numerical results support such conjecture.

\begin{figure}[!htbp]
  \centering
  \includegraphics[width=0.43\textwidth]{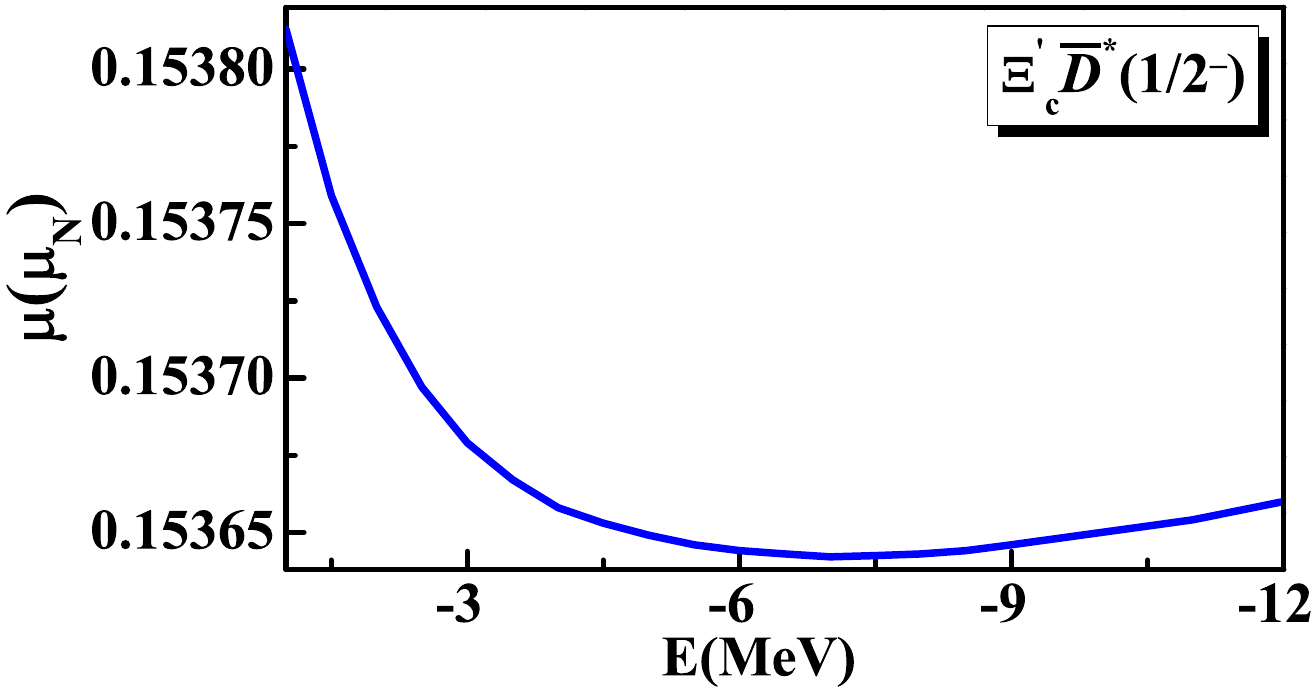}
  \includegraphics[width=0.43\textwidth]{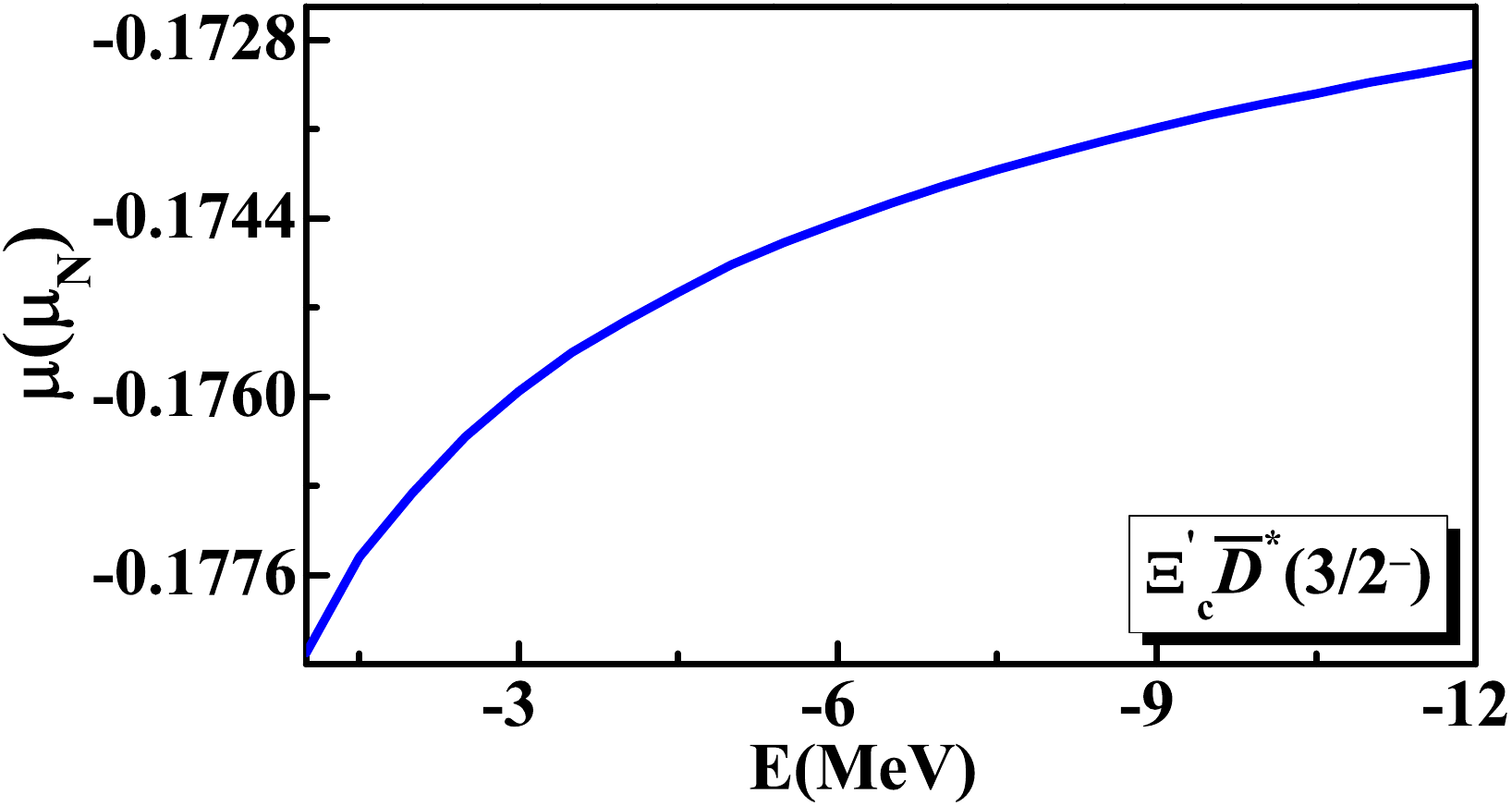}
  \caption{The magnetic moments dependent on the binding energies for the isoscalar $\Xi_c^{\prime}\bar D^{*}$ molecules with $J^P=1/2^-$ and $J^P=3/2^-$  after considering the $S$-$D$ wave mixing effect.}\label{SDMP}
\end{figure}

Due to the lack of the experimental data for the binding energies of the $S$-wave isoscalar $\Xi_c^{\prime}\bar D^{*}$ molecules with $J^P=1/2^-$ and $J^P=3/2^-$, while the hadronic molecule is a loosely bound state \cite{Chen:2016qju}, we discuss their magnetic moments by varying the binding energies of the $S$-wave isoscalar $\Xi_c^{\prime}\bar D^{*}$ molecular states with $J^P=1/2^-$ and $J^P=3/2^-$  in the range of $-1 \sim -12$ MeV \cite{Wang:2022mxy}. In Fig. \ref{SDMP}, we present the magnetic moment dependence of the binding energies for the isoscalar $\Xi_c^{\prime}\bar D^{*}$ molecular states with $J^P=1/2^-$ and $J^P=3/2^-$ by performing the $S$-$D$ wave mixing analysis.

After including the contribution of the $D$-wave channels, we can find that the magnetic moments of the isoscalar $\Xi_c^{\prime}\bar D^{*}$ molecular states with $J^P=1/2^-$ and $J^P=3/2^-$ are $0.1538~\mu_N\sim0.1536~\mu_N$ and $-0.1783~\mu_N\sim-0.1730~\mu_N$ with the binding energies of the corresponding systems in the range from $-1$ to $-12$ MeV, respectively. Furthermore, their magnetic moments are not obviously dependent on the binding energies of the $S$-wave isoscalar $\Xi_c^{\prime}\bar D^{*}$ molecules with $J^P=1/2^-$ and $J^P=3/2^-$. By comparing the numerical results for the single channel case, it is obvious that the $S$-$D$ wave mixing effect plays a rather minor role, and the change of their magnetic moments is less than $0.02\mu_N$ when adding the contribution of the $D$-wave channels since the components of the $D$-wave channels are very tiny \cite{Wang:2022mxy}.

\begin{figure}[!htbp]
  \centering
  \includegraphics[width=0.40\textwidth]{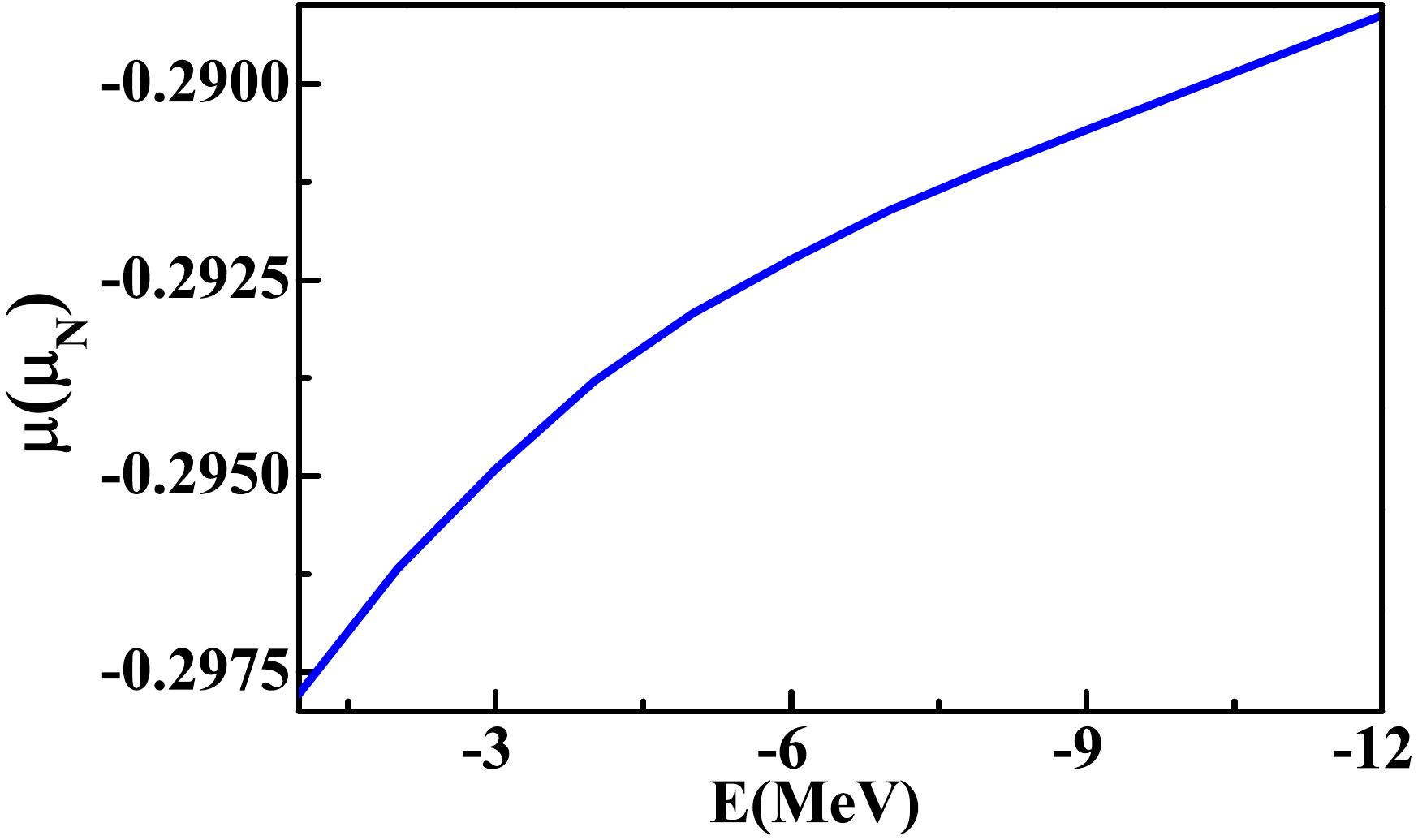}
  \caption{The transition magnetic moment of the $\Xi_c^{\prime}\bar D^{*}|{3}/{2}^-\rangle \to \Xi_c^{\prime}\bar D^{*}|{1}/{2}^-\rangle \gamma$ process dependent on the binding energies of the isoscalar $\Xi_c^{\prime}\bar D^{*}$ molecular states with $J^P=(1/2^-,\,3/2^-)$ after considering the $S$-$D$ wave mixing effect.}\label{SDTMP}
\end{figure}

Following the procedure discussed above, we also study the transition magnetic moment of the $\Xi_c^{\prime}\bar D^{*}|{3}/{2}^-\rangle  \to \Xi_c^{\prime}\bar D^{*}|{1}/{2}^-\rangle \gamma$ process by considering the $S$-$D$ wave mixing effect, and the relevant numerical result is given in Fig. \ref{SDTMP}. Here, we discuss the transition magnetic moment of the $\Xi_c^{\prime}\bar D^{*}|{3}/{2}^-\rangle \to \Xi_c^{\prime}\bar D^{*}|{1}/{2}^-\rangle \gamma$ process by scanning the binding energies of the $S$-wave isoscalar $\Xi_c^{\prime}\bar D^{*}$ molecular states with $J^P=(1/2^-,\,3/2^-)$  in the range of $-1 \sim -12$ MeV.

As presented in Fig. \ref{SDTMP}, the transition magnetic moment of the $\Xi_c^{\prime}\bar D^{*}|{3}/{2}^-\rangle  \to \Xi_c^{\prime}\bar D^{*}|{1}/{2}^-\rangle \gamma$ process is $-0.298~\mu_N\sim-0.289~\mu_N$ by tuning the binding energies of the corresponding discussed systems from $-1$ MeV to $-12$ MeV, which is slightly bigger than the numerical result $-0.305~\mu_N$ from the single channel analysis. On the other hand, it shows that the radiative decay width of the $\Xi_c^{\prime}\bar D^{*}|{3}/{2}^-\rangle  \to \Xi_c^{\prime}\bar D^{*}|{1}/{2}^-\rangle \gamma$ process becomes smaller when adding the contribution of the $D$-wave channels.

\section{Summary}\label{sec4}

In the past few years, 
the community has made big progress on exploring the hidden-charm molecular pentaquarks  \cite{Richard:2016eis,Lebed:2016hpi,Brambilla:2019esw,Liu:2019zoy,Chen:2022asf,Olsen:2017bmm,Guo:2017jvc,Meng:2022ozq}. Especially, the observaton of three $P_c$ states in 2019 \cite{Aaij:2019vzc} provides a strong evidence of the existence of the hidden-charm molecular pentaquark states in the hadron spectroscopy \cite{Li:2014gra,Karliner:2015ina,Wu:2010jy,Wang:2011rga,Yang:2011wz,Wu:2012md,Chen:2015loa}. Recently, the reported evidence of the $P_{cs}(4459)$  \cite{LHCb:2020jpq} and the observation of the $P_{\psi s}^{\Lambda}(4338)$ \cite{Pcs} may provide us a good opportunity to identify the hidden-charm molecular pentaquarks with strangeness. In Refs. \cite{Karliner:2022erb,Wang:2022mxy}, 
the characteristic mass spectrum of 
$S$-wave $\Xi_c^{(\prime)}\bar D^{(*)}$-type hidden-charm molecular pentaquarks with strangeness was proposed, which is stimulated by the $P_{cs}(4459)$  \cite{LHCb:2020jpq} and  $P_{\psi s}^{\Lambda}(4338)$ \cite{Pcs}. However, other properties of the $S$-wave isoscalar $\Xi_c^{(\prime)}\bar D^{(*)}$ molecular states are still waiting to be further discussed, which can provide further information to reflect the inner structures of these discussed $S$-wave isoscalar $\Xi_c^{(\prime)}\bar D^{(*)}$ molecular pentaquarks.

In this work, we study the electromagnetic properties of the $S$-wave isoscalar $\Xi_c^{(\prime)}\bar D^{(*)}$ molecular states through the constituent quark model. In the concrete calculation, we first focus on the magnetic moments of the $S$-wave isoscalar $\Xi_c^{(\prime)}\bar D^{(*)}$ molecular states. Our numerical results show that the magnetic moment of the $S$-wave isoscalar $\Xi_c^{(\prime)}\bar D^{*}$ molecule with $J^P=1/2^-$ is obviously different from that of the $S$-wave isoscalar $\Xi_c^{(\prime)}\bar D^{*}$ molecule with $J^P=3/2^-$. Thus, measuring the magnetic moments of them is an effective approach to reflect their spin-parity quantum numbers and their underlying structures.  After that, we extend our theoretical framework to study the transition magnetic moments between the $S$-wave isoscalar $\Xi_c^{(\prime)}\bar D^{(*)}$ molecular states. And then, we find several relations of  the transition magnetic moments. Especially, the transition magnetic moments of the $S$-wave isoscalar $\Xi_c^{(\prime)}\bar D^{*}$ molecules with $J^P=1/2^-$ and $J^P=3/2^-$ are different, which can provide us the critical information to identify the spin-parity quantum numbers of the $S$-wave isoscalar $\Xi_c^{(\prime)}\bar D^{*}$ molecular states. After obtaining the transition magnetic moments between the $S$-wave isoscalar $\Xi_c^{(\prime)}\bar D^{(*)}$ molecular states, we also discuss the radiative decay behaviors between the $S$-wave isoscalar $\Xi_c^{(\prime)}\bar D^{(*)}$ molecular states, where most of
the radiative decay widths between the $S$-wave isoscalar $\Xi_c^{(\prime)}\bar D^{(*)}$ molecules are around 5.0 keV.
Finally, we discuss the electromagnetic properties of the isoscalar $\Xi_c^{(\prime)}\bar D^{(*)}$ molecular states by considering the $S$-$D$ wave mixing effect \cite{Wang:2022mxy}. When making comparison of the numerical results with and without the $S$-$D$ wave mixing effect, we find that the $S$-$D$ wave mixing effect plays a rather minor role for the magnetic moments of the isoscalar $\Xi_c^{\prime}\bar D^{*}$ molecular states with $J^P=1/2^-$ and $J^P=3/2^-$ and the transition magnetic moment of the $\Xi_c^{\prime}\bar D^{*}|{3}/{2}^-\rangle  \to \Xi_c^{\prime}\bar D^{*}|{1}/{2}^-\rangle \gamma$ process, since the components of the $D$-wave channels are not dominant for these states \cite{Wang:2022mxy}.

In summary, although the hadronic molecular states have attracted much attention on both the theoretical and experimental sides \cite{Liu:2013waa,Hosaka:2016pey,Chen:2016qju,Richard:2016eis,Lebed:2016hpi,Brambilla:2019esw,Liu:2019zoy,Chen:2022asf,Olsen:2017bmm,Guo:2017jvc,Meng:2022ozq}, investigation of the electromagnetic properties of the hadronic molecules has not received plenty of attention.
In this work, we are dedicated to this issue. 
We hope that the present work can stimulate further experimental and theoretical
exploration of the electromagnetic properties of the hadronic molecular states.

\section*{ACKNOWLEDGMENTS}

This work is supported by the China National Funds for Distinguished Young Scientists under Grant No. 11825503, National Key Research and Development Program of China under Contract No. 2020YFA0406400, the 111 Project under Grant No. B20063, and the National Natural Science Foundation of China under Grant Nos. 12175091, 11965016, and 12047501.

\end{document}